\documentclass[aps,eqsecnum,nofootinbib,superscriptaddress,showpacs]{revtex4}
\def\be{\begin{eqnarray}}
\def\ee{\end{eqnarray}}

\def\p{\partial}

\usepackage{amsmath,amsfonts,amssymb,bm}
\usepackage{color}

\usepackage{graphicx}
\graphicspath{{figures/}}

\begin{document}
\title{A holographic stress-energy tensor near the Cauchy horizon 
inside a rotating black hole}  
\date{\today} 

\begin{abstract}
We investigate a stress-energy tensor for a CFT at strong coupling 
inside a small five-dimensional rotating Myers-Perry black hole 
with equal angular momenta by using the holographic method. 
As a gravitational dual, we perturbatively construct a black droplet 
solution by applying the ``derivative expansion" method, generalizing 
the work of Haddad~\cite{Haddad2012}, and analytically compute the 
holographic stress-energy tensor for our solution. 
We find that the stress-energy tensor is finite at both the future and past 
outer (event) horizons, and that the energy density is negative just outside 
the event horizons due to the Hawking effect. 
Furthermore, we apply the holographic method to the question of quantum 
instability of the Cauchy horizon since, by construction, 
our black droplet solution also admits a Cauchy horizon inside. 
% 
% Expressing our solution in terms of the advanced null coordinates, 
% our solution is extended to the inside of the future event horizon. 
We analytically show that the null-null component of the holographic 
stress-energy tensor negatively diverges at the Cauchy horizon, 
suggesting that a singularity appears there, 
in favor of strong cosmic censorship. 
% and that the quantum negative energy rapidly inflates any objects falling into the singularity at strong coupling regime.      
\end{abstract}

\author{Akihiro Ishibashi}  
\email{akihiro@phys.kindai.ac.jp}
\affiliation{Department of Physics, Kindai University, Higashi-Osaka, 
577-8502, Japan}
\author{Kengo Maeda} 
\email{maeda302@sic.shibaura-it.ac.jp}
\affiliation{Faculty of Engineering,
Shibaura Institute of Technology, Saitama, 330-8570, Japan}
\author{Eric Mefford}
\email{mefford@physics.ucsb.edu}
\affiliation{Department of Physics, University of California, Santa Barbara, CA 93106, USA}

\maketitle
%%%%%%%%%%%%%%%%%%%%%%
\section{Introduction}
%%%%%%%%%%%%%%%%%%%%%%
One of the outstanding issues in gravity is understanding quantum effects in regions of large spacetime curvature.
Although energy densities in typical classical fields such as electromagnetic fields are always non-negative, there have been a number of studies that 
show the appearance of negative energy density 
when quantum field effects are taken into account. For instance, it was shown (see, e.g.~\cite{FPR1998}) that the energy 
density for an observer falling into a singularity negatively 
diverges for some physical vacuum state. On the Cauchy horizon deep inside a charged black hole, 
the stress-energy tensor was calculated for a two-dimensional massless scalar field model and the energy density 
diverges at the horizon~\cite{Hiscock1977}. Calculations of a conformal 
scalar field in Taub-NUT-type cosmologies show that the stress-energy tensor negatively diverges on the Cauchy horizon even though 
the curvature remains small~\cite{HiscockKonkowski1982}. 
However, most studies have been made for free massless scalar field 
models and little attention has been given to strongly interacting field models 
such as CFTs at strong coupling.     
    
The AdS/CFT duality~\cite{adscft} provides a powerful tool to investigate CFTs at strong coupling on a fixed 
curved background spacetime. According to the dictionary of the duality, a CFT at strong coupling on a fixed 
$d$-dimensional spacetime is dual to a gravitational theory in $d+1$-dimensional AdS spacetime with a timelike boundary conformal to 
the $d$-dimensional spacetime. 
Motivated by the investigation of Hawking radiation in a model of a CFT at 
strong coupling, two types of black hole solutions were constructed in asymptotically locally AdS spacetimes~\cite{Haddad2013, HMR2010-1, 
HMR2010-2, HMR2010-3, CDMS2011, FMS2012, FischettiMarolf2012, SantosWay2012, FischettiSantos2013, FiguerasTunyasuvunakool2013, Eric2016}. 
One solution is called a ``black funnel" in which  
there is a single connected horizon extending from the conformal boundary to an asymptotically planar horizon in the bulk, and it is dual to the 
thermal equilibrium Hartle-Hawking vacuum state of the boundary theory. The other is called a ``black droplet"  
solution in which the horizon is disconnected from the planar horizon in the bulk, and it is dual to the Unruh vacuum state.       
In these models, negative energy density is observed outside the event horizon due to the Hawking effect. However, these solutions are 
quite complicated and their construction has required numerical methods; 
% except the lower dimensional case~($d\le 3$) \cite{HMR2010-1}
hence, it is difficult to analyze general properties of the stress-energy tensor inside the boundary black hole. 
It is then desirable to have some analytically constructed solutions for 
a black funnel/droplet. 
Recently some attempts along this direction have been made by 
Haddad~\cite{Haddad2012}~\footnote{
Note that this expansion method is essentially the same as the one
developed for the "blackfold approach" in \cite{EHNO2009}.
See also e.g. \cite{Armas2013}, for further applications of this method.
}, who, using a derivative expansion method, has constructed a five-dimensional {\it static} black droplet solution 
and computed the holographic stress-energy tensor for the corresponding dual quantum field in the background of a four-dimensional static black hole 
background~\cite{Haddad2013}~(See also for the lower dimensional case~($d\le 3$) \cite{HMR2010-1,HMR2010-3,
FischettiMarolf2012}). It is clearly interesting to generalize the line of research~\cite{Haddad2012, Haddad2013} performed for the 
static vacuum case to more general cases. In particular, including rotations would drastically change the causal structure inside 
the resultant black funnel/droplet motivating a study of the holographic consequences of strong coupling in quantum fields 
near the inner~(Cauchy) horizon.      
  
In this paper, we construct a {\it rotating} black droplet solution by generalizing the work~\cite{Haddad2013}.  
In general, including rotation makes the relevant analysis significantly more complicated compared to the static case. For example, 
if one attempts to add a rotation to the model of ~\cite{Haddad2013} so that the corresponding boundary field lives in a 
four-dimensional Kerr black hole--which is already cohomogeneity-two, then one would have to construct a 
five-dimensional bulk black droplet by solving a cohomogeneity-three system. In order to avoid this technical difficulty, instead 
of trying to add a rotation to a five-dimensional black droplet, we attempt to construct a six-dimensional rotating black droplet 
solution dual to a five-dimensional field theory in the background of the rotating Myers-Perry black hole~\cite{MyersPerry} 
with equal angular momenta, which is known to be cohomogeneity-one. 
In this case, the derivative expansion method enables us to 
reduce the bulk field equations to a set of ordinary differential equations, thereby making it possible to compute--analytically 
and explicitly within our expansion framework--the holographic stress-energy tensor for a CFT at strong coupling and 
large $N$ inside the five-dimensional rotating black hole. In addition, since quantum field theories in odd-dimensions are
not well understood, it is of considerable interest to study the behavior of quantum fields in a five-dimensional spacetime. 
In fact, motivated from recent interests in five-dimensional conformal field theory~(see e.~g.,~\cite{FiguerasTunyasuvunakool2013} for 
references), the six-dimensional rotating black droplet solutions dual to the rotating 
Myers-Perry black hole spacetime with equal angular momenta on the boundary were numerically constructed and the 
holographic stress-energy tensor was derived in region outside the event horizon~\cite{FischettiSantos2013, FiguerasTunyasuvunakool2013}. 

Having two rotations, the rotating droplet solution admits not only an outer event horizon but also an inner (Cauchy) horizon. In this paper, 
we are primarily concerned with the properties of the holographic stress-energy tensor inside the outer horizon and in particular, investigating the quantum 
instability of the Cauchy horizon. 
We find that the null-null component of the stress-energy tensor diverges negatively near the Cauchy horizon, 
in agreement with the study of free massless scalar fields~\cite{Hiscock1977, HiscockKonkowski1982}.  
Our results suggest that the Cauchy horizon suffers from a quantum instability 
in favor of the strong cosmic censorship. 
As far as we know, this is the first example of applying the holographic method 
to study the Cauchy horizon instability due to quantum effects. 
We also find that negative energy appears just outside the outer horizon, describing particle creation by the Hawking effect. 
Nevertheless, there is no flux at infinity. This suggests that the dual phase corresponds to a transition from black funnels to 
black droplets, and that it is reminiscent of soft condensed matter systems representing a transition 
from a fluid-like behavior to rigid behavior, just like a ``jammed" state~\cite{FischettiSantos2013} (see also \cite{Eric2016}).

The paper is organized as follows. In next section, we describe our metric ansatz, derive the equations of motion, and construct a rotating 
black droplet solution in six-dimensions by using the derivative expansion method. In section \ref{sec:III}, we perform an analytic computation of
the holographic stress-energy tensor for a CFT at strong coupling 
and large $N$ inside the five-dimensional rotating black hole on the boundary. 
In section~\ref{sec:IV}, we numerically check our results analytically obtained in the previous sections. Section~\ref{sec:V} is devoted to summary and discussion.

%%%%%%%%%%%%%%%%%%%%%%%%%%%%%
\section{Derivative expansion method}
\label{II}
%%%%%%%%%%%%%%%%%%%%%%%%%%%%%
In this section, we derive the field equations following the derivative expansion method~\cite{Haddad2012} and 
investigate general properties of the solution. Our bulk field equations 
are the $6$-dimensional vacuum Einstein equations with negative cosmological 
constant,
%Our theory is given by the action 
%\begin{align}
%\label{action}
%S=\int d^6x \sqrt{-g}\left[R+\frac{20}{l^2}\right]  
%\end{align}
\begin{align}
\label{Einstein_6dim}
R_{\mu\nu} = - \frac{5}{L^2}g_{\mu\nu},  
\end{align} 
where $L$ is the AdS radius.  
We start with the following metric ansatz:   
\begin{align}
\label{ansatz_null0}
& d \bar{s}^2=\left[\frac{L^2}{z^2F(z)}-\frac{z^2}{L^2}\Biggl(\frac{rF'(z)}{2F(z)} \Biggr)^2 \right]dz^2
+\frac{z^2r\alpha(r,z)F'(z)}{L^2F(z)}\{\sqrt{F(z)}\,dvdz-drdz\} \nonumber \\
& \qquad \quad +\frac{z^2}{L^2}\Biggl[-F(z)\frac{f(r,z)}{h(r,z)}dv^2
+2\sqrt{\frac{F(z)}{h(r,z)}}\,dvdr +\frac{r^2}{4}(d\theta^2+\sin^2\theta d\phi^2)    \nonumber \\
& \qquad \qquad \qquad +r^2h(r,z)\Bigl(d\psi+\frac{\cos\theta}{2} d\phi-\sqrt{F(z)}\Omega(r,z) dv\Bigr)^2   \Biggr], \nonumber \\
& f(r,z)=\left(1-\frac{r_+^2(z)}{r^2}\right)\left(1-\frac{\kappa^2r_+^2(z)}{r^2} \right), \qquad 
h(r,z)=1+\frac{\kappa^2r_+^4(z)}{r^4}, \nonumber \\
& \Omega(r,z)=\frac{\kappa\sqrt{1+\kappa^2}\,r_+^3(z)}{h(r,z)r^4}, 
% \qquad g(r,z)=\frac{1}{h(r,z)}, 
\qquad F(z)=1-\frac{\mu^5}{z^5}, 
\end{align}
% 
% \begin{align}
% \label{ansatz_null0}
% & ds_{(0)}^2=\left[\frac{L^2}{z^2F(z)}-\frac{z^2}{L^2}\Biggl(\frac{rF'(z)}{2F(z)} \Biggr)^2 \right]dz^2
% +\frac{z^2r\alpha(r,z)F'(z)}{L^2F(z)}\{\sqrt{F(z)}\,dvdz-drdz\} \nonumber \\
% & +\frac{z^2}{L^2}\Biggl[-F(z)f(r,z)g(r,z)dv^2+2\sqrt{F(z)g(r,z)}\,dvdr +\frac{r^2}{4}(d\theta^2+\sin^2\theta d\phi^2)    \nonumber \\
% &+r^2h(r,z)\Bigl(d\psi+\frac{\cos\theta}{2} d\phi-\sqrt{F(z)}\Omega(r,z) dv\Bigr)^2   \Biggr], \nonumber \\
% & f(r,z)=\left(1-\frac{r_+^2(z)}{r^2}\right)\left(1-\frac{\kappa^2r_+^2(z)}{r^2} \right), \qquad 
% h(r,z)=1+\frac{\kappa^2r_+^4(z)}{r^4}, \nonumber \\
% & \Omega(r,z)=\frac{\kappa\sqrt{1+\kappa^2}\,r_+^3(z)}{h(r,z)r^4}, \qquad g(r,z)=\frac{1}{h(r,z)}, \qquad F(z)=1-\frac{\mu^5}{z^5}, 
% \end{align}
% 
where $\alpha$ is an unknown function of $r$ and $z$ determined later. In the limit $r_+\to 0$ and $\alpha\to -1$, 
this metric reduces to the familiar planar Schwarzschild-AdS spacetime with horizon radius $\mu$ after performing the coordinate 
transformation, $v=t+r/\sqrt{F(z)}$. 
Furthermore, the metric at each $z=\mbox{const.}$ hypersurface represents 
the cohomogeneity-one Myers-Perry black hole solution with equal angular 
momenta~\cite{MyersPerry}, where the outer and inner (Cauchy) horizons are located at $r=r_+$ and 
$r=\kappa r_+$~($0\le \kappa<1$), respectively. So, the metric~(\ref{ansatz_null0}) represents a rotating black string embedded in the background planar Schwarzschild-AdS spacetime in which the horizon is extended along $z$-direction. 
The metric~(\ref{ansatz_null0}) itself does not satisfy 
the Eqs.~(\ref{Einstein_6dim}), and must be corrected order by order 
in derivatives. To this end, we write the metric as 
\begin{align}
\label{whole_metric}
& ds^2=d\bar{s}^2+ds_{(\epsilon)}^2, \nonumber \\
& ds_{(\epsilon)}^2=\sum_{n=1}^\infty \epsilon^n h^{(n)}_{\mu\nu}(r)dx^\mu dx^\nu,  
\end{align}
where $\epsilon$ is the formal derivative expansion parameter defined below and $h^{(n)}_{\mu\nu}(r)$ is the $n$th correction of the 
metric determined by the Einstein equations~(\ref{Einstein_6dim}). 
The derivative expansions are valid only when the horizon radius of the string is much smaller than the other scales, 
\begin{align}
\label{parameter_range}
r_+\ll \mu\sim L.   
\end{align}
This implies that the background metric~(\ref{ansatz_null0}) changes very slowly along the $z$-direction compared with 
the radial scale $r_+$. Thus, the contributions of the first and second derivatives with respect to 
$z$-direction to the Einstein Eqs.~(\ref{Einstein_6dim}) are suppressed by a factor of $r_+/L$ and 
$(r_+/L)^2$~(or similarly, $r_+/\mu$ and $(r_+/\mu)^2$).

Following~Ref.~\cite{Haddad2012}, we shall expand the metric functions, %$z$, 
$F$, $r_+$, and $\alpha$ in a series of $z-z_c$ around an arbitrary value $z_c$ as  
\begin{align}
\label{expansion}
% & z=z_c+\epsilon(z-z_c) + \cdots, \nonumber \\
& g(r,z) =g_c + \epsilon g_1 (z-z_c) + \epsilon^2 g_2 (z-z_c)^2+\cdots, 
\end{align}
% \begin{align}
% \label{expansion}
% & z=z_c+\epsilon(z-z_c), \nonumber \\
% & r_+(z)=r_c+\epsilon(z-z_c)r_1+\epsilon^2(z-z_c)^2r_2+\cdots, \nonumber \\ 
% & F=F_c+\epsilon(z-z_c)F_1+\epsilon^2(z-z_c)^2F_2+\cdots, \nonumber \\
% & \alpha(r,z)=\alpha(r;z_c)+\epsilon (z-z_c)\alpha_1(r)+\epsilon^2(z-z_c)^2\alpha_2(r)+\cdots,   
% \end{align}
where $g(r,z)$ collectively denotes the metric functions such as $F$, $r_+$, and $\alpha$, 
and the expansion coefficients are $g_n :=\p_z^n g(z_c)/n!$. Note that the expansion coefficients are functions of 
only $r$, but $F_n$ and $r_n$, are independent of $r$.

%Here, $\epsilon$ is the formal expansion parameter in which the integer $n$ of $\epsilon^n$ in Eqs.~(\ref{expansion}) 
%represents the $n$th derivative expansion around $z=z_c$, and we fix $\epsilon=1$ at the end. 
So, the Einstein Eqs.~(\ref{Einstein_6dim}) are formally modified to 
\begin{align}
\label{modified_Einstein}
r_c^2\,R_{\mu\nu}=-5\epsilon^2\frac{r_c^2}{L^2} g_{\mu\nu},  
\end{align}
where $r_c=r_+(z_c)$. This implies that the effect of the cosmological constant appears at second order in the derivative expansion~(\ref{expansion}). 
Note that the derivative expansion parameter $\epsilon$ 
will be set to unity at the end of our calculations. 

%%%%%%%%%%%%%%%%%%%%%%%%%
\subsection{First order in derivatives}
%%%%%%%%%%%%%%%%%%%%%%%%%
Substituting Eqs.~(\ref{ansatz_null0}) and (\ref{whole_metric}) into 
Eqs.~(\ref{modified_Einstein}) %(\ref{Einstein_6dim}) 
one finds that 
the field Eqs.~(\ref{modified_Einstein}) %(\ref{Einstein_6dim}) 
are satisfied at first order by 
\begin{align}
h^{(1)}_{\mu\nu}(r)=0, 
\end{align}
provided that the following two equations with respect to $\alpha(r;z_c)$ 
\begin{align}
\label{Eq:alpha}
& \alpha'+\frac{\alpha}{r}+\frac{(6F_cr_1+F_1r_c)r^8+24\kappa^2r_1F_c\,r_c^4\, r^4+\kappa^4r_c^8(2F_cr_1-F_1r_c)}
{4F_1r_c\,r^3(r^4+\kappa^2r_c^4)^{3/2}}=0, \nonumber \\
& \alpha''+\left(\frac{1}{r}+\frac{4r^3}{r^4+\kappa^2r_c^4}  \right)\alpha'-\left(\frac{1}{r^2}-\frac{4r^2}{r^4+\kappa^2r_c^4}  \right)\alpha
\nonumber \\
& +\frac{3F_1r^8+4\kappa^2r_c^3(F_1\,r_c-r_1F_c)\, r^4+\kappa^4r_c^7(4F_cr_1+F_1r_c)}
{F_1(r^4+\kappa^2r_c^4)^{5/2}}=0
\end{align}
are satisfied. When 
\begin{align}
\label{b1-sol}
r_1=\frac{r_cF_1}{2F_c} 
\end{align}
is satisfied, the solution $\alpha$ satisfying both two equations~(\ref{Eq:alpha}) is given by 
\begin{align}
\label{alpha_sol}
\alpha(r;z_c)=-\frac{r^2}{\sqrt{r^4+\kappa^2r_c^4}}+\frac{C}{r}, 
\end{align}
where $C$ is an integration constant. We discard the integration constant $C$ because it can be eliminated by gauge 
transformation of $v\to v+C/\sqrt{F}$. In this case, $\kappa\to 0$ limit agrees with the non-rotating four-dimensional 
black string case~\cite{Haddad2012}.   

%%%%%%%%%%%%%%%%%%%
\subsection{Second order in derivatives}
%%%%%%%%%%%%%%%%%%%
At second order, $O(\epsilon^2)$, we make an ansatz for the non-zero perturbed 
metric $h^{(2)}_{\mu\nu}$ as 

\begin{align}
\label{def:h2}
&h^{(2)}_{\mu\nu}dx^\mu dx^\nu=2\gamma(r)\left(d\psi+\frac{\cos\theta}{2} d\phi\right)dv+h_{vv}(r)dv^2+2h_{vr}(r)dvdr \nonumber \\
&\qquad \qquad \quad 
  +h_{zz}(r)dz^2+\beta(r)\left(d\psi+\frac{\cos\theta}{2} d\phi\right)^2 . 
% +2\rho(r)(d\psi_++A_\phi d\phi) dr. 
\end{align} 

% We derive equations of motion for the metric functions above. 
% Note that in the derivation, we have used $\alpha$ given by 
% (\ref{alpha_sol}) with $r_c$ replaced by $r_+$. 
We derive equations of motion for the metric functions above 
by substituting Eqs.~(\ref{ansatz_null0}), (\ref{whole_metric}), 
and (\ref{def:h2}) into (\ref{modified_Einstein}) and also using 
% Note that in the derivation, we have used 
$\alpha$ given by 
(\ref{alpha_sol}) with $r_c$ replaced by $r_+$. 
The equation of motion for $h_{zz}$ is decoupled from the other variables as 
\begin{align}
&-\frac{L^2F_c^2\,(r^2-r_c^2)(r^2-\kappa^2 r_c^2)(r^4+\kappa^2r_c^4)^3z_c^{10}}{5r^2}h_{zz}'' \nonumber \\
&-\frac{L^2F_c^2\,z_c^{10}\{3r^4-r_c^2(1+\kappa^2)r^2-\kappa^2r_c^4\}(r^4+\kappa^2r_c^4)^3}{5r^3}h_{zz}'+{\cal P}(r)=0,   
\end{align}
where the source term ${\cal P}$ is explicitly given by Eq.~(\ref{def P}). 
%The general solution includes two integral constants. 
% By imposing the regularity on the horizon {\color{blue}$r=r_c$}, we obtain the following analytic solution, 
The general solution includes two integral constants, one of 
which is determined by imposing the regularity on the horizon $r=r_c$. Then, we obtain the following analytic solution, 
\begin{align}
\label{Eq_zz_G}
& h_{zz}=-\frac{5r^2}{4L^2}+\frac{25\mu^{10}}{4L^2F_c^2\,z_c^{10}}\left(r^2+\frac{\kappa^2r_c^4-r^4}{\sqrt{r^4+\kappa^2r_c^4}} \right)
+\frac{5}{4L^2F_c\,z_c^5}\Bigl\{(z_c^5+5\mu^5)r^2-6\mu^5\sqrt{r^4+\kappa^2r_c^4}\Bigr\} \nonumber \\
& \qquad \quad 
+\frac{15(1+\kappa^2)\, r_c^2\,\mu^5}{2L^2F_c\,z_c^5}\ln(r^2-\kappa^2r_c^2)+C, 
\end{align}  
where $C$ is the remaining integral constant. 
Hereafter, we discard this constant becuase it can be eliminated by making a gauge transformation~\cite{Haddad2012}. 
% 
% Note that $h_{zz}$ in principle contains, apart from $C$, another integration 
% constant $C'$. In (\ref{Eq_zz_G}), we have already discarded $C'$ 
% by imposing the regularity at the horizon. 
% 
% Furthermore, $\rho$ is a pure gauge, as $\rho$ does not appear in the second 
% order Einstein Eqs.~(\ref{modified_Einstein}). Hence, we set $\rho=0$.  
% 
We find that we can solve for $h_{vr}$ in terms of the other variables, 
so we need only solve three coupled second order differential equations for 
$\gamma(r)$, $\beta(r)$, and $h_{vv}$. 
\begin{align}
\label{Eq_vr_G}
& h_{vr}=\frac{\sqrt{F_c}\,(2r^2-r_c^2(1+\kappa^2))r^3}{4(3r^4-\kappa^2\,r_c^4)\sqrt{r^4+\kappa^2r_c^4}}\,\beta'
-\frac{\kappa \sqrt{1+\kappa^2}\,r_c^3\,r^3}{(6r^4-2\kappa^2\,r_c^4)\sqrt{r^4+\kappa^2r_c^4}}\,\gamma' \nonumber \\
&-\frac{(3r^4+\kappa^2\,r_c^4)r^3}{4\sqrt{F_c}\,(3r^4-\kappa^2\,r_c^4)\sqrt{r^4+\kappa^2r_c^4}}h'_{vv}
+\frac{\sqrt{F_c}\,r^2\{r_c^6\,\kappa^2(1+\kappa^2)+3r_c^4\,\kappa^2r^2+r_c^2(1+\kappa^2)r^4-r^6\}}
{2(3r^4-r_c^4\,\kappa^2)(r^4+r_c^4\,\kappa^2)^{3/2}}\beta \nonumber \\
&+\frac{\kappa\sqrt{1+\kappa^2}\,r_c^3\,r^2}{(3r^4-\kappa^2\,r_c^4)\sqrt{r^4+\kappa^2r_c^4}}\gamma
-\frac{r^2h_{vv}}{2\sqrt{F_c}\,\sqrt{r^4+\kappa^2r_c^4}}   
+\frac{F_c^{3/2}z_c^4\{\kappa^2r_c^4-2r_c^2(1+\kappa^2)r^2+3r^4\}r^3}{4L^4(3r^4-\kappa^2r_c^4)\sqrt{r^4+\kappa^2r_c^4}}h'_{zz}
\nonumber \\
&+\frac{5r^4}{8\sqrt{F_c}L^6z_c^6(3r^4-\kappa^2\,r_c^4)(r^4+\kappa^2r_c^4)^{5/2}}\times \nonumber \\
& \Biggl[8F_c^2z_c^{10} (r^4+\kappa^2r_c^4)^2r^4+5\{6r^{12}+2\kappa^4r_c^8\,r^4-2\kappa^6r_c^{12}+3\kappa^4(1+\kappa^2)r_c^{10}r^2  
\nonumber \\
&+2\kappa^2r_c^4((1+\kappa^2)r_c^2-\sqrt{r^4+\kappa^2r_c^4})r^6
-((1+\kappa^2)r_c^2+6\sqrt{r^4+\kappa^2r_c^4})r^{10} \nonumber \\
&+2r_c^2(5\kappa^2r_c^2+2(1+\kappa^2)\sqrt{r^4+\kappa^2r_c^4})r^8\}\mu^{10}-8F_c^2z_c^{10}(r^4+\kappa^2r_c^4)^2r^4\Biggr], 
\end{align}   
\begin{align}
\label{Eq_gamma_G}
& \sqrt{F_c}L^6r^2(r^2-r_c^2)(r^2-\kappa^2r_c^2)(3r^4-\kappa^2r_c^4)(r^4+\kappa^2r_c^4)^2\gamma'' \nonumber \\
&+\sqrt{F_c}L^6r(r^4+\kappa^2r_c^4)^2\{3r^8-3(1+\kappa^2)r_c^2\,r^6+2\kappa^2r_c^4\,r^4-7\kappa^2(1+\kappa^2)r_c^6\,r^2-r_c^8\kappa^4 \}\gamma'
\nonumber \\
&-4\sqrt{F_c}L^6(r^4+\kappa^2r_c^4)^3\{3r^4-3(1+\kappa^2)r_c^2\,r^2-\kappa^2r_c^4\}\gamma \nonumber \\
& +2F_c L^6 \kappa \sqrt{1+\kappa^2}\,r_c^3\,r(r^4+\kappa^2r_c^4)^2\{r^4-2(1+\kappa^2)r_c^2\,r^2+\kappa^2r_c^4\}\beta' \nonumber \\
&-8L^6\kappa^3 \sqrt{1+\kappa^2}\,r_c^7\,r^3(r^4+\kappa^2r_c^4)^2h'_{vv} \nonumber \\
&+4F_cL^6\kappa \sqrt{1+\kappa^2}\,r_c^3(r^4+\kappa^2r_c^4)^2\{r^4+2(1+\kappa^2)r_c^2\,r^2+\kappa^2r_c^4 \}\beta \nonumber \\
& +4F_c^2L^2z_c^4\kappa \sqrt{1+\kappa^2}\,r_c^3\,r^3(r^4+\kappa^2r_c^4)^2
\{3r^4-2(1+\kappa^2)r_c^2\,r^2+\kappa^2r_c^4 \}h'_{zz}+{\cal S}(r)=0, 
\end{align}
\begin{align}
\label{Eq_Beta_G}
& -F_cL^6r(r^2-r_c^2)(r^2-\kappa^2r_c^2)(3r^4-\kappa^2r_c^4)(r^4+\kappa^2r_c^4)^2\beta'' \nonumber \\
&+F_cL^6(r^4+\kappa^2r_c^4)^2\{3r^8-9(1+\kappa^2)r_c^2\,r^6+6\kappa^2r_c^4\,r^4-5\kappa^2(1+\kappa^2)r_c^6\,r^2+3r_c^8\kappa^4\}\beta' 
\nonumber \\
& +12F_cL^6r(r^4+\kappa^2r_c^4)^3\{2r^2+(1+\kappa^2)r_c^2 \}\beta \nonumber \\
&-8\sqrt{F_c}L^6\kappa\sqrt{1+\kappa^2}\,r_c^3\,r^2(3r^4+\kappa^2r_c^4)(r^4+\kappa^2r_c^4)^2\gamma' \nonumber \\
&-8L^6\kappa^2r_c^4\,r^2(3r^4+\kappa^2r_c^4)(r^4+\kappa^2r_c^4)^2h'_{vv} \nonumber \\
&+16\sqrt{F_c}L^6\kappa\sqrt{1+\kappa^2}\,r_c^3\,r(3r^4+\kappa^2r_c^4)(r^4+\kappa^2r_c^4)^2\gamma \nonumber \\
&+8F_c^2L^2\kappa^2z_c^4r_c^4\,r^2(r^4+\kappa^2r_c^4)^2\{3r^4-2(1+\kappa^2)r_c^2\,r^2+\kappa^2r_c^4 \}h'_{zz}+{\cal R}(r)=0, 
\end{align}
\begin{align}
\label{Eq_vv_G}
& -L^6r^2(r^2-r_c^2)(r^2-\kappa^2r_c^2)(3r^4-\kappa^2r_c^4)(r^4+\kappa^2r_c^4)^2h''_{vv} \nonumber \\
&-L^6r(r^4+\kappa^2r_c^4)^2\{9r^8-9(1+\kappa^2)r_c^2\,r^6-6\kappa^2r_c^4\,r^4+7\kappa^2(1+\kappa^2)r_c^6\,r^2+\kappa^4r_c^8 \}h'_{vv}
\nonumber \\
& -2F_cL^6(1+\kappa^2)r_c^2\,r^3(r^4+\kappa^2r_c^4)^2\{(1+\kappa^2)r_c^2-2r^2\}\beta' \nonumber \\
& +4\sqrt{F_c}L^6\kappa\sqrt{1+\kappa^2}r_c^3\,r(r^4+\kappa^2r_c^4)^2\{3r^4-(1+\kappa^2)r_c^2\,r^2-\kappa^2r_c^4\}(r\gamma'-2\gamma)
\nonumber \\
&-4F_cL^6(1+\kappa^2)r_c^2(r^4+\kappa^2r_c^4)^2\{4r^4-(1+\kappa^2)r_c^2\,r^2-2\kappa^2r_c^4\}\beta \nonumber \\
&+2F_c^2L^2(1+\kappa^2)r_c^2z_c^4\,r^3(r^4+\kappa^2r_c^4)^2\{3r^4-2(1+\kappa^2)r_c^2\,r^2+\kappa^2r_c^4\}h'_{zz}+{\cal Q}(r)=0, 
\end{align} 
where ${\cal S}(r)$, ${\cal R}(r)$, and ${\cal Q}(r)$ are functions of $r$ given by Eqs.~(\ref{def_S}), (\ref{def_R}), and (\ref{def_Q}) in the Appendix. 
From the other constraint equations, we obtain the coefficient $r_2$ as 
\begin{align}
\label{b2_sol}
r_2=\frac{r_c(4F_2F_c-F_1^2)}{8F_c^2}. 
\end{align}
Combining Eqs.~(\ref{b1-sol}) and (\ref{b2_sol}), we obtain 
\begin{align}
r_+(z)=r_0\sqrt{F(z)},  
\end{align} 
up to second order in the derivative expansion, where $r_0$ is the radius of $r_+$ at the AdS boundary, $z\to \infty$. Just as 
in the non-rotating five-dimensional black string case~\cite{Haddad2012}, the droplet horizon shrinks to zero at the horizon of the 
planar Schwarzschild-AdS spacetime, ending on the horizon.

These three equations (\ref{Eq_gamma_G}), (\ref{Eq_Beta_G}), and (\ref{Eq_vv_G}) have a singular source term $\sim (r-r_c\kappa)^{-1}$ 
arising from $h_{zz}$ in (\ref{Eq_zz_G}). This implies that $\gamma$, $\beta$, and $h_{vv}$ can be expanded near the inner (Cauchy) 
horizon as 
\begin{align}
\label{expansion_inner}
& \gamma(r)\simeq \ln(r-\kappa\,r_c)\{a_0+a_1(r-\kappa\,r_c)+a_2(r-\kappa\,r_c)^2+\cdots\}+d_0+d_1(r-\kappa\,r_c)+\cdots, \nonumber \\
& \beta(r)\simeq \ln(r-\kappa\,r_c)\{b_0+b_1(r-\kappa\,r_c)+b_2(r-\kappa\,r_c)^2+\cdots\}+e_0+e_1(r-\kappa\,r_c)+\cdots, \nonumber \\ 
& h_{vv}(r)\simeq \ln(r-\kappa\,r_c)\{c_0+c_1(r-\kappa\,r_c)+c_2(r-\kappa\,r_c)^2+\cdots\}+f_0+f_1(r-\kappa\,r_c)+\cdots. \nonumber \\
% & \gamma(r)=a\ln(r-\kappa\,r_c)+a_1+a_2(r-r_c)+\cdots, \nonumber \\
% & \beta(r)=b\ln(r-\kappa\,r_c)+b_1+b_2(r-r_c)+\cdots, \nonumber \\
% & h_{vv}(r)=c\ln(r-\kappa\,r_c)+c_1+c_2(r-r_c)+\cdots \nonumber \\
\end{align}
%with only one constraint equation among the coefficients $a_i$, $b_i$, $c_i$~($i=1,\,2$).   
% Cosistency of these Eqs.~(\ref{Eq_gamma_G}), (\ref{Eq_Beta_G}), and (\ref{Eq_vv_G}) at $r=\kappa r_c$ tells us that  
% 
Note that we have assumed that the black droplet solution is non-extremal, i.~e.~, $\kappa<1$, in the expansion. 
 Substituting these into Eqs.~(\ref{Eq_gamma_G}), (\ref{Eq_Beta_G}), and (\ref{Eq_vv_G}), we obtain all the coefficients 
provided that the coefficients $c_0$, $d_0$, $e_0$, $f_0$, $e_1$, and $f_1$ are given. This implies that 
six independent mode solutions exist for the second order differential equations~(\ref{Eq_gamma_G}), (\ref{Eq_Beta_G}), and 
(\ref{Eq_vv_G}). For the discussions in the next section, it suffices to obtain the relation between the leading order coefficients 
$a_0$, $b_0$, and $c_0$. The remaining subleading coefficients are determined by numerics in Sec.~\ref{sec:IV}. 

The leading coefficients $a_0$ and $b_0$ are determined by $c_0$ as  
% 
% Substituting these into Eqs.~(\ref{Eq_gamma_G}), (\ref{Eq_Beta_G}), and (\ref{Eq_vv_G}), {\color{green}we obtain all the coefficients 
% provided that the leading coefficients $c_0$, $a_1$, $b_1$, $d_0$, $e_0$, and $f_0$ are given. This implies that 
% six-independent mode solutions exist for the second order differential equations~(\ref{Eq_gamma_G}), (\ref{Eq_Beta_G}), and 
% (\ref{Eq_vv_G})}. The other leading coefficient $a_0$ and $b_0$ are determined by   
\begin{align}
\label{coefficient_CH}
& a_0=-\frac{r_c\,\kappa\sqrt{1+\kappa^2}\{2L^6z_c\,c_0+15r_c^2(1+\kappa^2)\mu^5F_c\}}{2L^6z_c(1-\kappa^2)\sqrt{F_c}}, \nonumber \\
& b_0=\frac{r_c^2\,\kappa^2\{2L^6z_c\,(1+3\kappa^2)c_0+15r_c^2(3+4\kappa^2+\kappa^4)\mu^5F_c\}}{2L^6(1-\kappa^2)z_c\,F_c}. 
\end{align}

% \begin{align}
% \label{coefficient_CH}
% a=-\frac{45\sqrt{F_c}\,r_c^3\kappa(1+\kappa^2)^{3/2}\mu^5}{L^6z_c(1-9\kappa^2)}, \nonumber \\
% b=\frac{60\,r_c^4\,\kappa^2(1+\kappa^2)\mu^5}{L^6z_c(1-9\kappa^2)}, \nonumber \\  
% c=\frac{15F_cr_c^2(5+8\kappa^2+3\kappa^4)\mu^5}{2L^6z_c(1-9\kappa^2)}.   
% \end{align}

By Eq.~(\ref{Eq_vr_G}), we also find the asymptotic behavior of $h_{vr}$ near the Cauchy horizon:
\begin{align}
\label{vr_Asy}
h_{vr}\simeq \frac{r_c\,\kappa^2\{2L^6z_c\,\kappa^2c_0+15(1+\kappa^2)r_c^2\,\mu^5F_c\}}
{4L^6z_c(1-\kappa^2)\sqrt{1+\kappa^2}\sqrt{F_c}(r-\kappa r_c)}. 
\end{align}

% By Eq.~(\ref{Eq_vr_G}), we also find the asymptotic behavior of $h_{vr}$ near the Cauchy horizon:
% \begin{align}
% \label{vr_Asy}
% h_{vr}\simeq \frac{15\mu^5\sqrt{F_c}\,r_c^3\,\kappa^2(1-3\kappa^2)\sqrt{1+\kappa^2}}{4L^6z_c(1-9\kappa^2)(r-\kappa r_c)}. 
% \end{align}
% The five free parameters among the coefficients $a_i$, $b_i$, $c_i$~($i=1,\,2$) are determined by both the boundary 
% condition at the outer horizon $r=r_c$ and at the null infinity, $r=\infty$. Since we are interested in the Unruh vacuum, we shall 
% impose the regularity condition at the outer horizon. Note that the singular behavior near the inner horizon is determined 
% by the logarithmic terms in Eqs.~(\ref{expansion_inner}), which is independent of the boundary conditions.     

%%%%%%%%%%%%%%%%%%%%%%%%%%%%%%%%%%%
\subsection{The non-rotating case}
%%%%%%%%%%%%%%%%%%%%%%%%%%%%%%%%%%%
In the non-rotating case~($\kappa=0$), 
Eqs.~(\ref{Eq_gamma_G}) and (\ref{Eq_Beta_G}) respectively
% Eqs.~(\ref{Eq_Beta_G}) and (\ref{Eq_vv_G}) 
for $\gamma$ and $\beta$ are decoupled from the 
other variables and we can set $\gamma=\beta=0$. 
Furthermore, we obtain analytic expressions for $h_{vv}$ and $h_{vr}$ 
from Eqs.~(\ref{Eq_zz_G}), (\ref{Eq_vv_G}), and (\ref{Eq_vr_G}):  
\begin{align}
\label{kappa_0_sol}
& h_{zz}=\frac{15r_c^2\,\mu^5\ln r}{L^2F_c\,z_c^5}, \nonumber \\
& h_{vv}=C_2-\frac{C_1}{2r^2}-\frac{5\mu^5F_c\,r_c^2}{L^6z_c\,r^2}
\{r_c^2-(r^2-2r_c^2)\ln r\}, \nonumber \\
& h_{vr}=\frac{-4C_2L^6z_c^6+5r_c^2\,\mu^5(4z_c^5+\mu^5)-20r_c^2\,\mu^5z_c^5\,F_c \ln r}{8L^6z_c^6\sqrt{F_c}},  
\end{align}
where $C_1$ and $C_2$ are constants that correspond to a global shift in 
the temperature as explained in \cite{Haddad2012}, so we must set it to zero.  
%Note also that in the expression of $h_{zz}$ above, we have already discarded 
%an integration constant by making a gauge transformation. 

%%%%%%%%%%%%%%%%%%%%%%%%%%%%%%%%%%%%%
\section{The holographic stress-energy tensor}
\label{sec:III}
%%%%%%%%%%%%%%%%%%%%%%%%%%%%%%%%%%%%%
In this section, we calculate the holographic stress-energy tensor using the prescription of \cite{EJM1999}, up to 
the second order in $\epsilon$. In the six-dimensional bulk theory, the regularized action becomes 
\begin{align}
\label{regularized_action}
& S = \frac{1}{16\pi G_6} \int_{{\cal M}} dx^6\sqrt{-g} 
                                       \left(R+\frac{30}{L^2}\right) 
      +\frac{1}{8\pi G_6}\int_{\p {\cal M}} dx^5 \sqrt{-q} K  
\nonumber \\
      & \qquad +\frac{1}{8\pi G_6}\int_{\p {\cal M}} dx^5 \sqrt{-q} 
                          \left[ \frac{4}{L}+\frac{L}{6}{\cal R}
                                +\frac{L^3}{18}\left({\cal R}_{ab}{\cal R}^{ab}
                                -\frac{5}{16}{\cal R}^2\right)+\cdots 
                          \right], 
\end{align}
where ${\cal R}$ is the Ricci scalar of the induced metric $q_{ab}=g_{ab}-n_a n_b$ at $z=z_c$ associated with the unit normal outward pointing vector $n^a$, 
and $K$ is the trace of the extrinsic curvature defined below. 
Note that the first three terms 
in the second line are sufficient to cancel the divergences. 
Furthermore, the last two terms are at $O(\epsilon^4)$, 
since the induced metric is the vacuum Myers-Perry black hole~\cite{MyersPerry} at zeroth order, i.~e.~, 
${\cal R}_{ab}={\cal R}=O(\epsilon^2)$. Thus, 
the holographic stress-energy tensor $T_{ab}$, given by $T_{ab}=(2/\sqrt{-q})\,\delta S/\delta q^{ab}$, 
becomes 
\begin{align}
\label{energy_momentum}
T_{ab}=\frac{L}{8\pi G_{6}}\left[\frac{1}{3}E_{ab}-\frac{\epsilon}{L}\Bigl(K_{ab}-q_{ab}K \Bigr)-\frac{4\epsilon^2}{L^2}q_{ab}   \right]
+O(\epsilon^4), 
\end{align}
where $E_{ab}$ is the Einstein tensor of the induced metric $q_{ab}$, and $K_{ab}$ is the extrinsic curvature defined by 
\begin{align}
K_{ab}={q_a}^c \nabla_c n_b. 
\end{align}
If the metric (\ref{whole_metric}) is decomposed into 
\begin{align}
ds^2=(N^2+N_a N^a)dz^2+2N_a dx^a dz+q_{ab}dx^a dx^b, 
\end{align}
$K_{ab}$ is rewritten by 
\begin{align}
K_{ab}=\frac{1}{2N}(\p_z q_{ab}-D_a N_b-D_b N_a), 
\end{align}
where $D_a$ is the covariant derivative with respect to the induced metric $q_{ab}$, and the lapse function $N$ and the shift vector 
$N_a$ are given by  
\begin{align} 
& N_v=\frac{5\alpha r\mu^5}{2L^2z_c^4}+O(z_c^{-9}), \quad 
N_r=-\frac{5\alpha r\mu^5 }{2L^2z_c^4}+O(z_c^{-9}), \quad \mbox{the other components}=0, 
\nonumber \\
& N=\frac{L}{z_c\sqrt{F_c}}+O(\epsilon^2). 
\end{align} 
Note that $N_a=O(\epsilon)$, as it includes the derivative with respect to $z$ from Eq.~(\ref{ansatz_null0}). Thus, 
if we expand $q_{ab}$, $K_{ab}$, and $E_{ab}$ as 
\begin{align}
& q_{ab}={q^{(0)}}_{ab}+\epsilon^2{q^{(2)}}_{ab}+\cdots, \nonumber \\
& K_{ab}= \epsilon K^{(1)}_{ab}+\epsilon^3 K^{(3)}_{ab}+\cdots, \nonumber \\
& E_{ab}=\epsilon^2E^{(2)}_{ab}+\cdots, 
\end{align}
$K^{(1)}_{ab}$ is determined by ${q^{(0)}}_{ab}$ as 
\begin{align}
K^{(1)}_{ab}=\frac{z_c\sqrt{F_c}}{2L}(\p_z {q^{(0)}}_{ab}-\bar{D}_a N_b-\bar{D}_b N_a), 
\end{align} 
where $\bar{D}_a$ denotes the covariant derivative with respect to ${q^{(0)}}_{ab}$. 
Then, Eq.~(\ref{energy_momentum}) reduces to 
\begin{align}
\label{energy_momentum1}
T_{ab}=\frac{\epsilon^2 L}{8\pi G_{6}}\left[\frac{1}{3}E^{(2)}_{ab}-\frac{1}{L}\Bigl(K^{(1)}_{ab}-q^{(0)}_{ab}K^{(1)} \Bigr)
-\frac{4}{L^2}q^{(0)}_{ab}   \right]+O(\epsilon^4).  
\end{align}
This implies that the second order perturbation $h^{(2)}_{\mu\nu}$ contributes to the stress-energy tensor only through 
the Einstein tensor, up to $O(\epsilon^2)$. 

First, we investigate the stress-energy tensor in the static case~($\kappa=0$). Substitution of Eqs.~(\ref{kappa_0_sol}) into 
Eq.~(\ref{energy_momentum1}) yields 
\begin{align}
\label{Stress_Energy_static}
& T_{vv}= \epsilon^2 \cdot C \cdot 
        \frac{4r^6-9r_c^2\,r^4+5r_c^6}{r^6}, \nonumber \\
& T_{vr}= \epsilon^2 \cdot C \cdot 
        \frac{-4r^4+5r_c^2\,r^2+5r_c^4}{r^4}, \nonumber \\
& T_{rr}= \epsilon^2 \cdot C \cdot 
        \frac{5(r^2-r_c^2)}{r^2}, \nonumber \\
& T_{\psi\psi}=\frac{2}{\cos \theta} T_{\psi\phi}= \epsilon^2 \cdot C \cdot 
        \frac{r^4-5r_c^4}{r^2}, \nonumber \\
% & T_{\psi\phi}= C \cdot \epsilon^2 \cdot 
%              \frac{r^4-5r_c^4}{2r^2}\cos\theta, \nonumber \\
& T_{\theta\theta}= T_{\phi\phi} = \epsilon^2 \cdot C \cdot 
             \frac{r^4-5r_c^4}{4r^2}, \nonumber \\
% & T_{\phi\phi}= C \cdot \epsilon^2 \cdot 
%                \frac{r^4-5r_c^4}{4r^2},   \nonumber \\
%%% 
% & 8\pi G_6T_{vv}=\frac{\epsilon^2(4r^6-9r_c^2\,r^4+5r_c^6)\mu^5}{2L^3z_c^3\,r^6}, \nonumber \\
% & 8\pi G_6T_{vr}=\frac{\epsilon^2(-4r^4+5r_c^2\,r^2+5r_c^4)\mu^5}{2L^3z_c^3\,r^4}, \nonumber \\
% & 8\pi G_6 T_{rr}=\frac{5\epsilon^2(r^2-r_c^2)\mu^5}{2L^3z_c^3\,r^2}, \nonumber \\
% & 8\pi G_6 T_{\psi\psi}=\frac{\epsilon^2(r^4-5r_c^4)\mu^5}{2L^3z_c^3\,r^2}, \nonumber \\
% &  8\pi G_6 T_{\psi\phi}=\frac{\epsilon^2(r^4-5r_c^4)\mu^5}{4L^3z_c^3\,r^2}\cos\theta, \nonumber \\
% & 8\pi G_6 T_{\theta\theta}=\frac{\epsilon^2(r^4-5r_c^4)\mu^5}{8L^3z_c^3\,r^2}, \nonumber \\
% & 8\pi G_6 T_{\phi\phi}=\frac{\epsilon^2(r^4-5r_c^4)\mu^5}{8L^3z_c^3\,r^2}.  \nonumber \\
\end{align}   
where $C=\mu^5/16 \pi G_6 L^3z_c^3$. 
It is easily checked that the conservation law $\bar{D}_a T^{ab}=0$ is satisfied. Near the outer horizon $r=r_c$, negative energy density appears, 
i.e.,~$T_{vv}<0~(r>r_c)$. % , due to the Hawking effect. 
This implies that due to the Hawking effect,
pair creation of particles occurs near the horizon, 
and the negative energy particles are absorbed into the horizon. 
Nevertheless, there is no flux at null infinity. This is verified 
by checking that the $(t,r)$-component of the stress-energy tensor 
in the original coordinate system $(t,r)$ becomes zero at null infinity. 
This is due to strong coupling effects of the dual CFT in the boundary 
theory, just as in the five-dimensional case~\cite{Haddad2013}. 
It is also immediately checked that the trace of our stress-energy tensor 
vanishes, in agreement with the general argument that odd dimensional CFTs 
have a vanishing trace anomaly.

Next, we investigate the stress-energy tensor near the inner~(Cauchy) horizon 
in the rotating case. 
Note that $K_{ab}$ is regular near the Cauchy horizon $r=\kappa r_c$ 
at $O(\epsilon)$ because ${q^{(0)}}_{ab}$ and the shift vector $N_a$ are 
regular there. 
Thus, the dominant term of $T_{ab}$ in Eq.~(\ref{energy_momentum}) near the Cauchy horizon comes from the Einstein 
tensor $E_{ab}$. As shown in Eqs.~(\ref{expansion_inner}), (\ref{coefficient_CH}) and (\ref{vr_Asy}), the second order metric $h_{ab}$ diverges 
near the Cauchy horizon. So, the relevant (i.e., $(r,r)$-) component of 
the Einstein tensor $E_{ab}$ can be expanded as 
% \begin{align}
% \label{E_vv}
% E_{vv}=\frac{\epsilon^2}{z_c^3}
%       \left[ \frac{C_1}{r-\kappa r_c}
%      +C_2\ln (r-\kappa r_c)+\cdots \right]+O(\epsilon^4), 
% \end{align}   
\begin{align}
 \label{E_rr}
 E_{rr}=\frac{\epsilon^2}{z_c^3}
 \left[-\frac{15r_c^2(1+\kappa^2)\mu^5}{4L^4(r-r_c\kappa)^2} 
  + \frac{C'}{r-\kappa r_c}+\cdots\right]+O(\epsilon^4), 
\end{align} 
%  \begin{align}
%  \label{E_vr}
%  E_{vr}=\frac{\epsilon^2}{z_c^3}\left[\frac{C_4}{r-\kappa r_c}
%       + C_5\ln (r-\kappa r_c)+\cdots \right]+O(\epsilon^4), 
%  \end{align}   
%  \begin{align}
%  \label{E_pp}
%  E_{\psi\psi}=\frac{\epsilon^2}{z_c^3}\left[\frac{C_6}{r-\kappa r_c}
%    + C_7\ln (r-\kappa r_c)+\cdots \right]+O(\epsilon^4), 
%  \end{align} 
%  \begin{align}
%  \label{E_tt}
%  E_{\theta\theta}=\frac{\epsilon^2}{z_c^3}\left[\frac{C_8}{r-\kappa r_c}
%  +C_{9}\ln (r-\kappa r_c)+\cdots \right]+O(\epsilon^4), 
%  \end{align}  
where $C'$ is a constant. As for the other components, 
the leading term in order $O(\epsilon^2)$ behave as $1/(r-\kappa r_c)$, 
and therefore are irrelevant to the rest of our arguments. 
%$C_i~(i=1,2,\cdots)$ are constants. 

The most striking feature is that $E_{rr}$ in Eq.~(\ref{E_rr}) negatively diverges at the Cauchy horizon. 
This implies that the null energy condition is strongly violated along the null direction, $\p_r$ near the Cauchy horizon: 
\begin{align}
\label{Trr_CH}
T_{rr}\simeq -\frac{5\epsilon^2r_c^2(1+\kappa^2)\mu^5}{32\pi G_6L^3z_c^3(r-r_c\kappa)^2}\to -\infty. 
\end{align}  
Interestingly, this behavior is very similar to the case of massless scalar 
field in two-dimensions \cite{Hiscock1977,HiscockKonkowski1982};  
in both cases, the stress-energy tensor negatively diverges as 
$(r- \kappa r_c)^{-2}$.

%%%%%%%%%%%%%%%%%%%%%%%
\section{Numerical results}
\label{sec:IV}
%%%%%%%%%%%%%%%%%%%%%%%

When we add rotation to our droplets, we must solve the second order equations numerically. To account for the logarithmic 
divergences in $\beta, \gamma$ and $h_{vv}$, as well as the pole in $h_{vr}$, we make the following ansatz,
\begin{align}
\label{numericalansatz}
\beta(r) &= \beta_L(r)\ln(r-\kappa r_c)+\beta_1(r)  \nonumber \\
\gamma(r) &= \gamma_L(r)\ln(r-\kappa r_c) + \gamma_1(r)  \nonumber \\
h_{vv}(r) &=h_{vvL}(r)\ln(r-\kappa r_c) + h_{vv1}(r)  \nonumber \\
h_{vr}(r) &=h_{vrL}(r)\ln(r-\kappa r_c) + \frac{r}{r-\kappa r_c}h_{vr1}(r).
\end{align}

\begin{figure}[t]
\includegraphics[scale=.5]{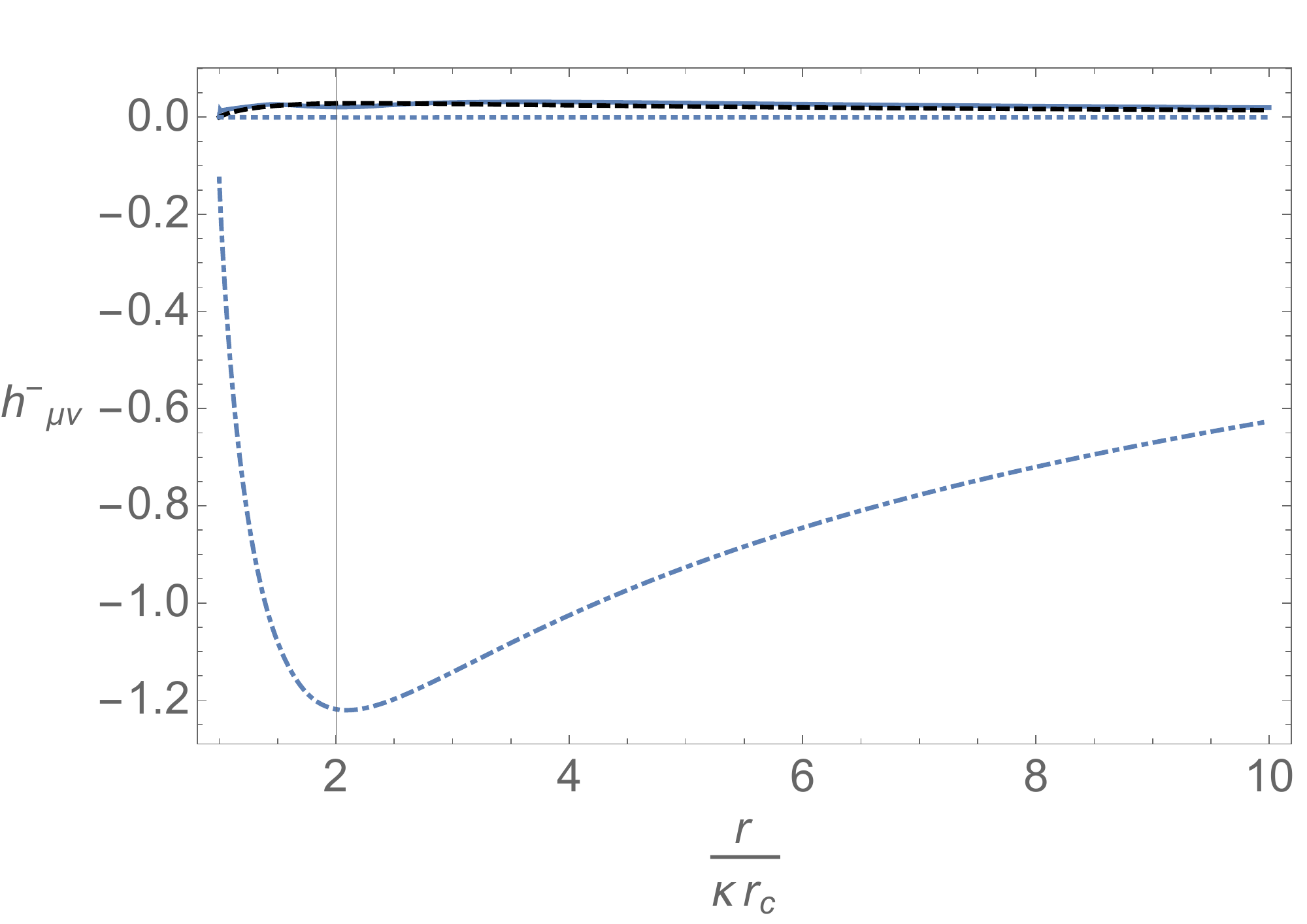}
\caption{\label{numsols} Here we plot the finite pieces of $h_{\mu\nu}(r)$ at $\mathcal{O}(\epsilon^2)$. The curves correspond to $\beta^{-}(r)$ (dotted), $\gamma^{-}(r)$ (dashed, black), $h_{vv}^{-}(r)$ (dot-dashed), and $h_{vr}^{-}(r)$ (thick) for $r_c=.1,\;\mu=1,\;\kappa=.2,\;z_c=(4\mu)^{1/5}$. Note that $\gamma^-$ and $h_{vr}^-$ nearly overlap, and $\beta^{-}$ is much smaller than the other functions.}
\end{figure}

In the last equation, the coefficient of $h_{vr1}(r)$ is required to have $h_{vv1}(r)$ vanish as $r$ goes to infinity, 
matching the non-rotating case.

 We insert these ansatz into (\ref{Eq_vr_G}) and find eight equations to solve 
numerically--four from the coefficients of $\ln(r-\kappa r_c)$ involving only $\beta_L, \gamma_L, h_{vvL}$ and $h_{vrL}$ 
and four remaining equations involving these variables as well as $\beta_1, \gamma_1, h_{vv1}, h_{vr1}$. It is numerically convenient
 to also set $L=1$ and work in terms of a variable $R\equiv 1/r$ in order to impose boundary conditions at spatial infinity. 
 
 As we did in the analytic case, we can perform a series expansion in powers of $R-(\kappa r_c)^{-1}$ near the Cauchy horizon to find appropriate boundary conditions on our new metric functions,
 \begin{align}
 X(R) = \sum_{i=0}^\infty x_i(R-\frac{1}{\kappa r_c})^{i}
 \end{align}
 where $X$ refers collectively to $\{\beta_L, \gamma_L, h_{vvL}, h_{vrL},\beta_1, \gamma_1, h_{vv1}, h_{vr1}\}$. This expansion reflects the fact that the divergences in $\beta,\gamma, h_{vv}$ come only from a log term sourced by $h_{zz}$ and there is an extra divergence of $(r-\kappa r_c)^{-1}$ in $h_{vr}$. Inserting this expansion into our eight differential equations and solving order by order in $(R-\frac{1}{\kappa r_c})$ leads to the following boundary conditions,
\begin{align}
h_{{vrL}}\left(R=\frac{1}{\kappa  r_c}\right)&= -\frac{\kappa  \left(15 \left(\kappa ^2+1\right) \left(2 \kappa ^6+7 \kappa ^4+12 \kappa ^2+3\right) \mu ^5 F_c r_c^2+2 \kappa ^2 \left(5 \kappa ^4+14 \kappa ^2+5\right) z_c h_{{vvL}}\left(\frac{1}{\kappa  r_c}\right)\right)}{4 \left(\kappa ^2-1\right)^2 \left(\kappa ^2+1\right)^{3/2} \sqrt{F_c} z_c}  \nonumber  \\
h_{{vvL}}'\left(\frac{1}{\kappa  r_c}\right)&= -\frac{\kappa  \left(\kappa ^2+1\right) r_c \left(15 \left(\kappa ^2+1\right) \mu ^5 F_c r_c^2+2 z_c h_{{vvL}}\left(\frac{1}{\kappa  r_c}\right)\right)}{\left(\kappa ^2-1\right) z_c}
\end{align}
as well as the previously derived conditions, Eq. (\ref{coefficient_CH}). Furthermore, the expansion leads to the following constraint at the horizon,
\begin{align}
0=&16 \kappa  \left(\kappa ^2-1\right) r_c^2 z_c \left(2 \left(\kappa ^2+1\right) \sqrt{F_c} \gamma _1\left(\frac{1}{\kappa  r_c}\right)+\sqrt{\kappa ^2+1} h_{{vv1}}'\left(\frac{1}{\kappa  r_c}\right)\right) \nonumber \\
&+5 \kappa ^2 \mu ^5 r_c^5 \biggl(3 \left(4 \kappa ^5-8 \kappa ^3+93 \sqrt{\kappa ^2+1} \kappa ^2+33 \sqrt{\kappa ^2+1}+19 \sqrt{\kappa ^2+1} \kappa ^6+63 \sqrt{\kappa ^2+1} \kappa ^4+4 \kappa \right) F_c \nonumber \\
&-20 \kappa  \left(\kappa ^2-1\right) \left(\kappa ^2-2 \sqrt{\kappa ^2+1} \kappa -1\right)\biggr)-4 \left(\sqrt{\frac{1}{\kappa ^2}+1}-\kappa ^3 \sqrt{\kappa ^2+1}\right) \nonumber \\
&-8 \sqrt{F_c} r_c z_c \left(3 \kappa  \left(\sqrt{\frac{1}{\kappa ^2}+1}-\kappa ^3 \sqrt{\kappa ^2+1}\right) \sqrt{F_c} \beta _1\left(\frac{1}{\kappa  r_c}\right)-2 \left(\kappa ^4-1\right) \gamma _1'\left(\frac{1}{\kappa  r_c}\right)\right) \nonumber \\
&F_c z_c \beta _1'\left(\frac{1}{\kappa  r_c}\right)+2 \kappa ^2 \sqrt{\kappa ^2+1} \left(57 \kappa ^4+48 \kappa ^2-1\right) r_c^3 z_c h_{{vvL}}\left(\frac{1}{\kappa  r_c}\right).
\end{align}
We can likewise perform a series expansion at spatial infinity in powers of $R$ (recall $R=0$ corresponds to spatial infinity) to find appropriate boundary conditions. This leads to
\begin{align}
\beta_L\left(0\right) &= \gamma_L(0) = 0, &h_{vr1}(0) = \frac{-5 \left(\kappa ^2+1\right) \mu ^5 \left(F_c-5\right) r_c^2-4 z_c h_{{vv1}}(0)}{8 \sqrt{F_c} z_c}  \nonumber, \\
h_{vvL}(0) &= \frac{5 \left(\kappa ^2+1\right) \mu ^5 F_c r_c^2}{z_c}, \quad &h_{vrL}(0) = -\frac{5 \left(\kappa ^2+1\right) \mu ^5 \sqrt{F_c} r_c^2}{2 z_c}  \nonumber, \\
\beta_1(0) &=\frac{25 \kappa ^2 \mu ^{10} r_c^4}{2 F_c z_c^6},\quad &\gamma_1(0) = -\frac{25 \kappa  \sqrt{\kappa ^2+1} \mu ^{10} r_c^3}{2 \sqrt{F_c} z_c^6}.
\end{align}
Note that these boundary conditions correspond to imposing 
a single constraint on the free parameters $c_0, d_0, f_0, e_1$ and $f_1$ in 
Eq.~(\ref{expansion_inner}). 
\begin{figure}[t]
\includegraphics[scale=.43]{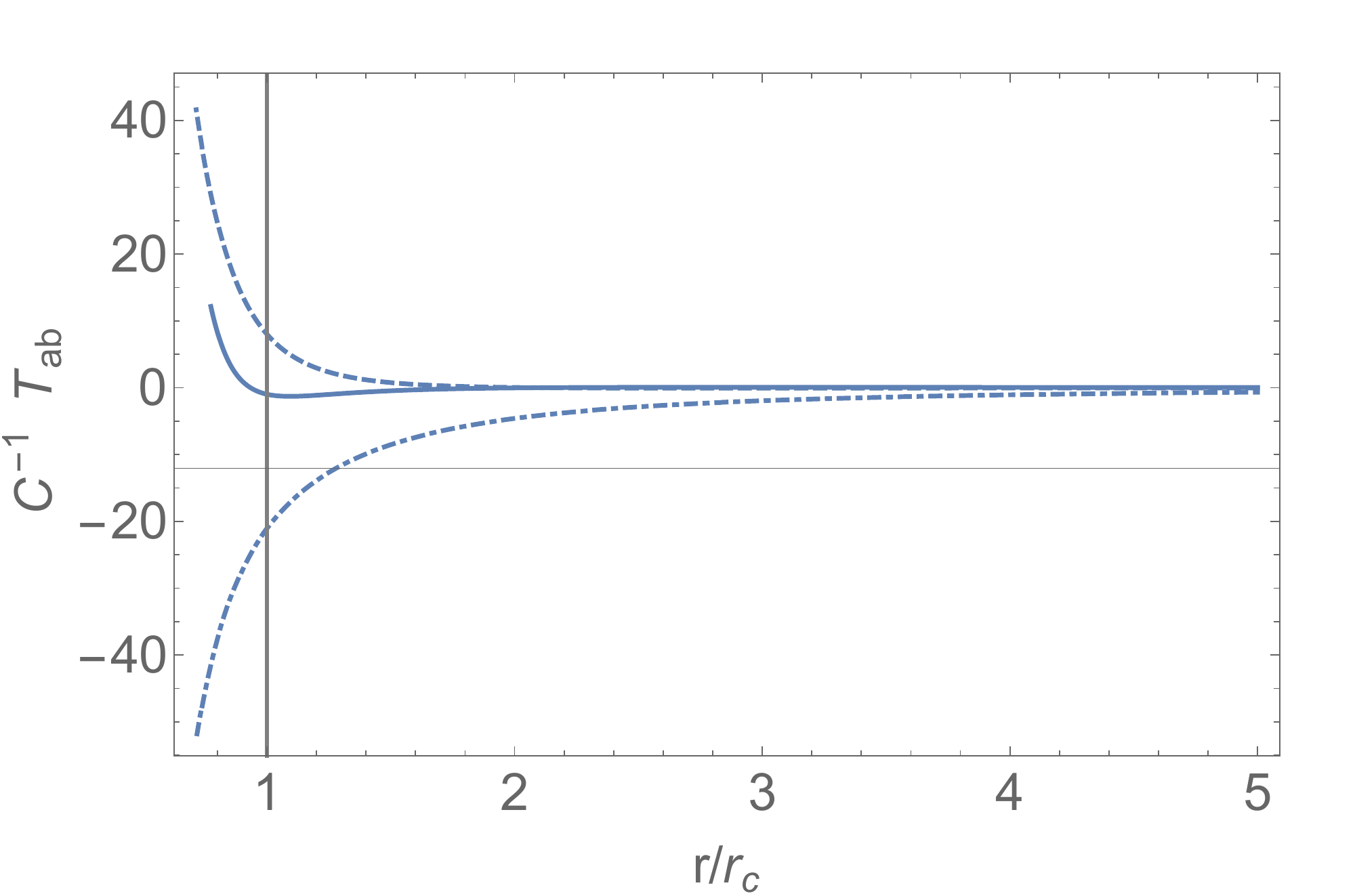}
\includegraphics[scale=.43]{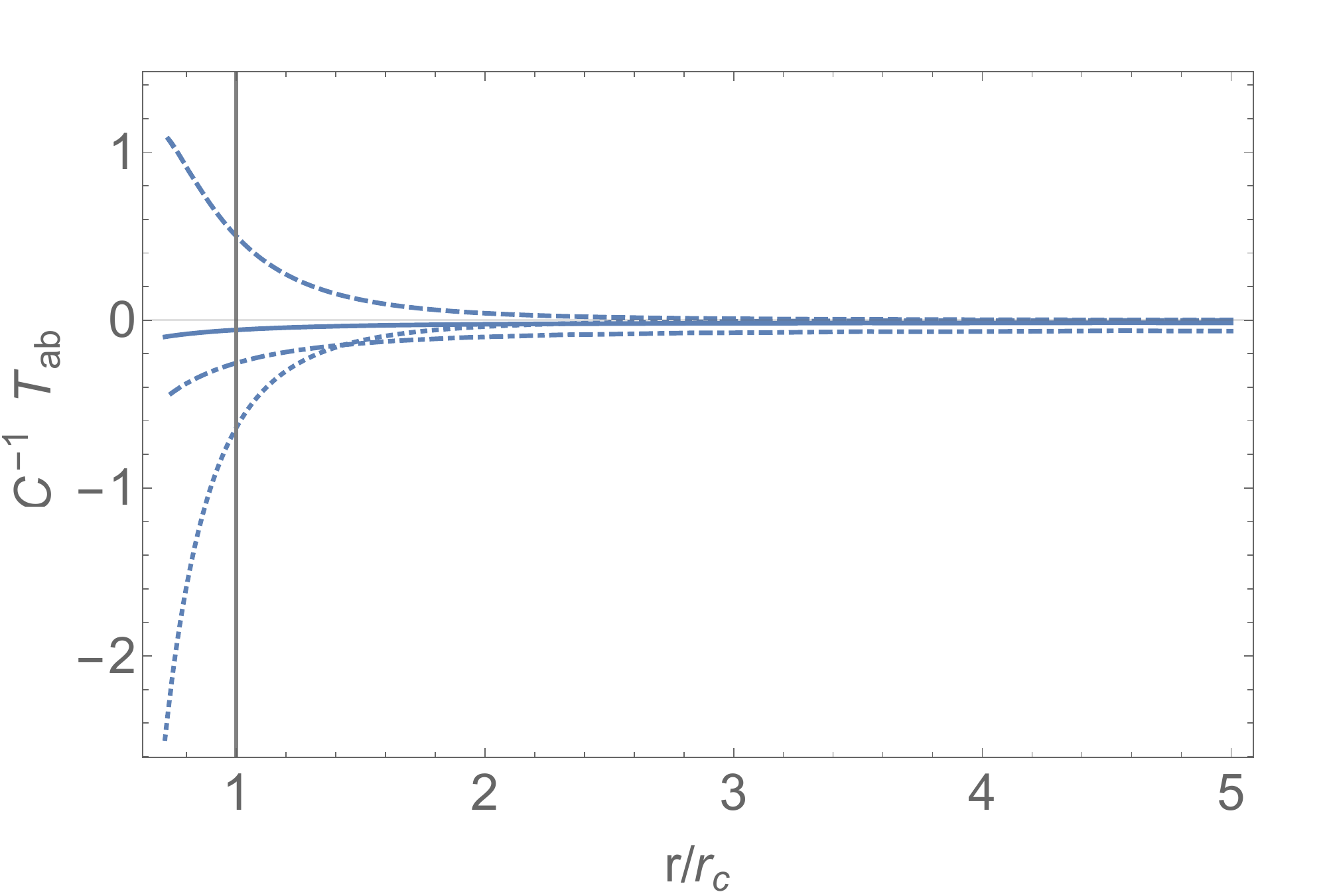}
\caption{\label{numstress} Here we plot the components of the holographic stress-energy tensor for the same parameters as figure \ref{numsols} with $\theta=\pi/2$ and $z_c = (10^8\mu)^{1/5}$. In the left plot, we show $T_{vr}$ (dashed), $T_{rr}$ (dot-dashed), and $T_{vv}$ (thick). In the right, we show $T_{\psi r}$ (dotted), $T_{\psi v}$ (dashed), $T_{\psi\psi}$ (dot-dashed), and $T_{\theta\theta}$ (thick). For each of these, we have scaled our solution by $C^{-1}$ defined below Eq.~(\ref{Stress_Energy_static}). Notably, each of these components is regular at the outer horizon (labelled by the vertical line at $r/r_c =1$).}
\end{figure}
Finally, there are a few boundary conditions which me must impose by hand. These are analogous to the constants $C_1$ and $C_2$ in Eq. (\ref{kappa_0_sol}). To smoothly match onto the non-rotating case, we choose $\beta_1(\frac{1}{\kappa r_c}) = \gamma_1(\frac{1}{\kappa r_c})=0$ and $h_{vv1}(0) = 0$.
This choice is equivalent to imposing $d_0=e_0= f_0=0$.

These boundary 
conditions are not sufficient to ensure smooth solutions because the point $R=3^{1/4}/(r_c\sqrt{\kappa})$ is a 
(regular) singular point of our differential equations. To accommodate this singularity, we used two numerical 
regions, $0\leq R \leq \frac{3^{1/4}}{r_c\sqrt{\kappa}}$ and $\frac{3^{1/4}}{r_c\sqrt{\kappa}}<R<\frac{1}{\kappa r_c}$ (this is only necessary for $\kappa<\sqrt{3}$). 
We impose continuity of our functions and match the first derivatives of our functions at this point. Regularity of the differential equation, 
or similarly smoothness of $h_{vrL}$ and $h_{vr1}$ at our singular point, amounts to two 
constraints. In total, we start with four free constants, $\beta_1(\frac{1}{\kappa r_c}),\gamma_1(\frac{1}{\kappa r_c}), h_{vvL}(\frac{1}{\kappa r_c})$, and $h_{vv1}(0)$ and fix three by hand to smoothly match onto the non-rotating solution. The final constant is fixed by consistency of the two constraints coming from the smoothness of $h_{vr1}, h_{vrL}$.

To find these numerical solutions, we use the Newton-Raphson method with pseudospectral collocation over a Chebyshev grid in the two numerical domains. 
In figure \ref{numsols}, we have plotted our solutions for $r_c = .1, \mu = 1, \kappa =.2, z_c = (4\mu)^{1/5}$ (reexpressed in terms of the original radial coordinate $r$). Importantly,
we have included only the finite pieces of the solutions, subtracting off the divergent pieces. For example, using the notation of (\ref{expansion_inner}),
\begin{align}
\label{reghmn}
\beta^{-}(r) \equiv \beta(r) - b_0\ln(r-\kappa r_c), \quad \gamma^{-} \equiv \gamma^{-}(r) -a_0\ln(r-\kappa r_c), \quad h_{vv}^{-}(r) = h_{vv}(r) - c_0\ln(r-\kappa r_c)
\end{align} 
and similarly for $h_{vr}^{-}(r)$. 

We have also plotted the non-vanishing components of the stress-energy tensor for this solution in figure \ref{numstress}. We have only included the part of the stress-energy tensor near $r=r_c$ because
the behavior of the stress-energy tensor near the Cauchy horizon can be derived from (\ref{coefficient_CH}) as was done for $T_{rr}$ in (\ref{Trr_CH}).
To verify that we obtained the correct holographic stress-energy tensor, we varied $z_c$ between $(10^4\mu)^{1/5}$ and $(10^8\mu)^{1/5}$ and checked that $C^{-1}( T_{ab})$ did not change.
 
\begin{figure}[t]
\includegraphics[scale=.47]{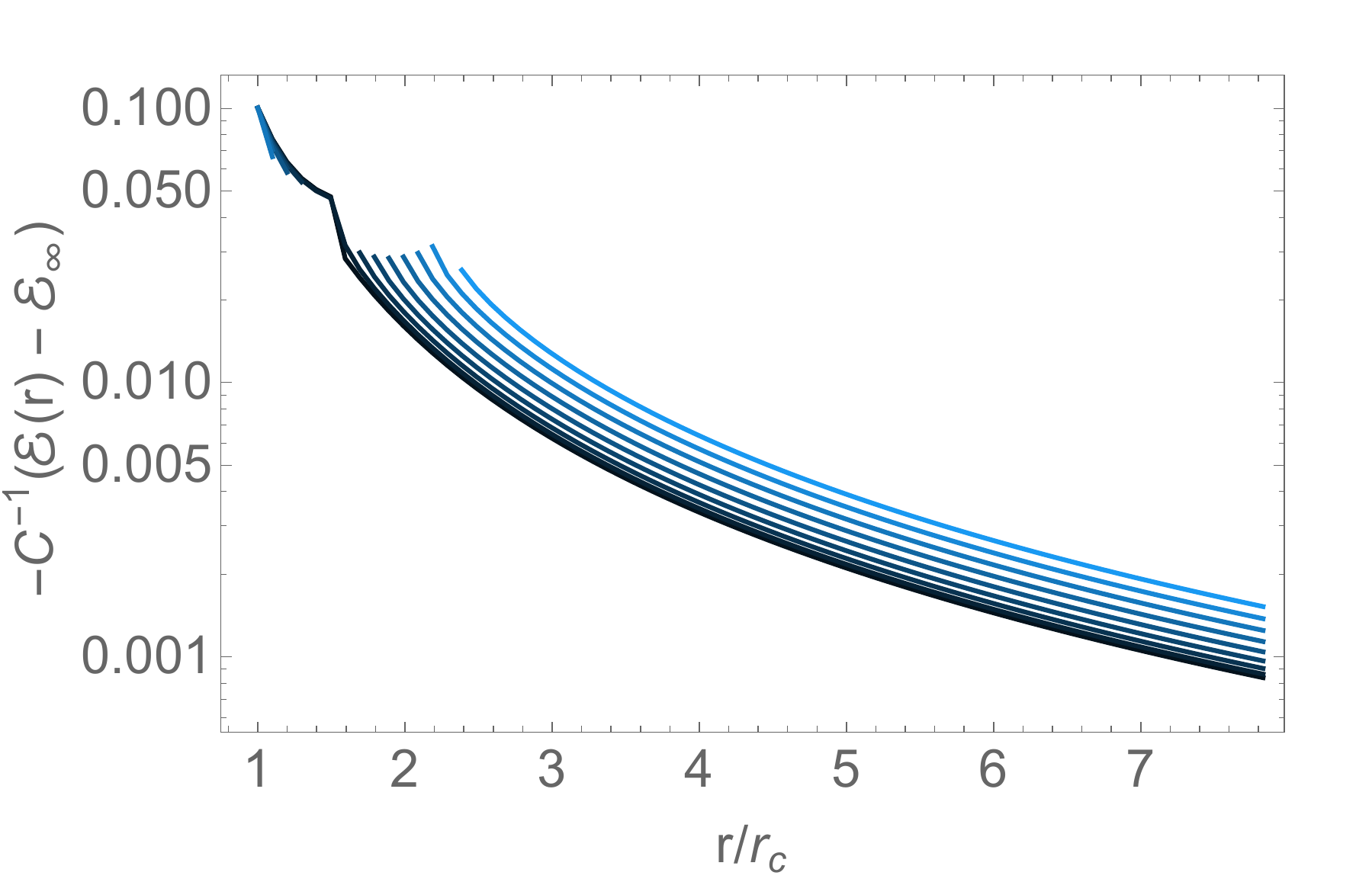}
\includegraphics[scale=.47]{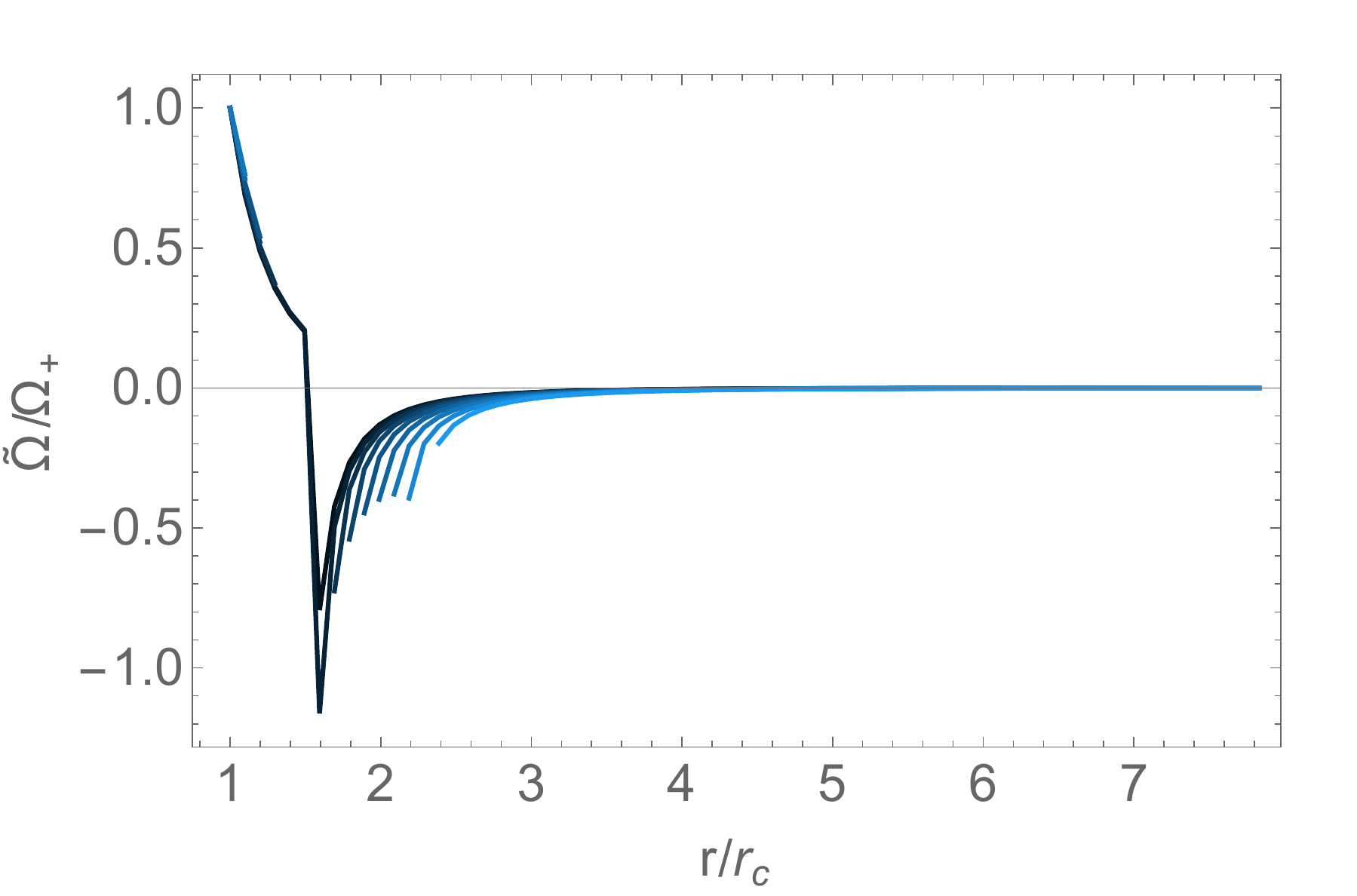}
\caption{\label{Tvv_numeric} (Left) Here we plot $\mathcal{E}(r)$ near the outer horizon for $\kappa = .1,\;.2,\;.3,\;.4,\;.5,\;.6,\;.7,\;.8,\;.9$ with $r_c=.1, \mu=1, z_c=(10^8\mu)^{1/5}$. The color of the curves gets lighter as $\kappa$ increases. As before, we have scaled the energy density by $C^{-1}$ and set $\theta=\pi/2$. The upper left of the figure corresponds to the near horizon region. The curves are discontinuous because there is a region set by $\kappa$ outside the event horizon (Eq.~\ref{complexeigen}), where the stress-energy tensor is not diagonalizable. (Right) Here we plot the rotation $\tilde{\Omega}(r)$ for the same values of $\kappa$. At the outer horizon, the rotation matches the value $\Omega(r_c,z_c)$ showing that $t$ is the generator for the outer horizon. As before, discontinuities arise because the stress-energy tensor is not diagonalizable.}
\end{figure}

As pointed out for the non-rotating case, an interesting quantity is the energy density near the outer horizon. The local energy density may be found by diagonalizing the stress-energy tensor $(T^{a}_{\;\;b})$, as done in \cite{FiguerasTunyasuvunakool2013}. The stress-energy tensor in our spacetime is diagonalizable near the horizon and far from the horizon, but there is an intermediate region
\begin{align}
\label{complexeigen}
r_c\sqrt{1-\kappa+\kappa^2+(1-\kappa)\sqrt{1+\kappa^2}}<r<r_c\sqrt{1+\kappa+\kappa^2+(1+\kappa)\sqrt{1+\kappa^2}}
\end{align}
 where the stress-energy tensor diagonalization breaks down. This is likely a result of our expansion, as in a fully non-perturbative solution like \cite{FiguerasTunyasuvunakool2013}, no such region was seen, though it is notable that our solution contains a finite temperature, rather than extremal, bulk horizon. Following \cite{FiguerasTunyasuvunakool2013}, In the region where this decomposition is well-defined, we may write
\begin{align}
T^{a}_{\;\;b}t^b=-\mathcal{E}(r)t^a
\end{align}
where $t^a$ is the (unique) normalized timelike eigenvector and $\mathcal{E}(r)$ can be interpreted as the energy density observed by the timelike observer with velocity $t^a$. At leading order in $r$ and $z_c$,
\begin{align}
t= \frac{1}{z_c}\left[\left(1-2\frac{(1+\kappa^2)r_c^2}{r^2}\pm\frac{\kappa^2 r_c^4}{r^4}\right)\frac{\partial}{\partial v} -\frac{\kappa\sqrt{1+\kappa^2}r_c^3}{r^4}\frac{\partial}{\partial\psi}\right] + \mathcal{O}\left(\frac{1}{r^5},\frac{\mu^5}{z_c^5}\right).
\end{align}
The plus sign in this equation corresponds to the near horizon region, while the minus sign corresponds to the region far from the horizon.

The energy density obtained from the decomposition is plotted in figure \ref{Tvv_numeric} for different values of $\kappa$. Just as in the non-rotating case, the region of negative energy density extends all the way from the horizon to spatial infinity. Interestingly, at spatial infinity, the energy density approaches a constant,
\begin{align}
\mathcal{E}_\infty\equiv\lim_{r\to\infty} \mathcal{E}(r) = \epsilon^2\cdot\frac{4C}{z_c^2} + \mathcal{O}(r^{-2}).
\end{align}
This should not be surprising because far from the boundary black hole, the CFT should be in a thermal state, with an energy density corresponding to the temperature of the bulk black hole. In fact, this value matches the energy density for a CFT dual to a 6 dimensional planar-AdS Schwarzschild black brane. Furthermore, this value is independent of $\kappa$ as it should be, since our boundary black holes are asymptotically flat and a similar result was seen for $\kappa=0$ in \cite{Haddad2013}. In figure \ref{Tvv_numeric}, we have subtracted this asymptotic value from the energy density to emphasize that a local observer near the black hole measures an energy density less than the thermal energy density because of quantum effects in the curved background spacetime. 

Interestingly, our energy density approaches $\mathcal{E}_\infty$ as $r^{-2}$, rather than the $r^{-7}$ decay observed in \cite{FiguerasTunyasuvunakool2013}. This less steep fall-off could be a consequence of our derivative expansion method. However, it is also notable that our droplet solution ends on a finite temperature black brane horizon, whereas in \cite{FiguerasTunyasuvunakool2013}, the bulk horizon was extremal (the Poincar\'e horizon) and the black droplet was disconnected. Similar fall-off discrepancies were seen in numerical constructions of five dimensional static droplets, where the energy density decayed as $r^{-5}$ with an extremal bulk horizon \cite{Figueras2011}, but as $r^{-1}$ for a finite temperature bulk horizon \cite{SantosWay2014}.  Importantly, as in the analytic case, for an observer with tangent vector $t^a$, for all choices of $\kappa$, the energy density diverges negatively as $(r- \kappa r_c)^{-2}$ near the Cauchy horizon.

We also can use the stress-energy tensor eigenvalue decomposition to define rotation of the dual plasma. Again, following \cite{FiguerasTunyasuvunakool2013}), we write the timelike eigenvector of the stress-energy tensor as
\begin{align}
T = \frac{\partial}{\partial v} + \tilde{\Omega}(r,z)\frac{\partial}{\partial \psi}
\end{align}
and define $\tilde{\Omega}$ to be the rotation. At the outer horizon, this becomes (at zeroth order in $\epsilon$)
\begin{align}
T_+ = \frac{\partial}{\partial v} + \sqrt{F(z_c)}\Omega(r_c,z_c)\frac{\partial}{\partial \psi}
\end{align}
which, on the conformal boundary, matches the future generator of the horizon at $r=r_c$. Note that the rotation decays as $r^{-4}$, rather than the $r^{-2}$ fall-off seen in \cite{FiguerasTunyasuvunakool2013}. The faster fall-off could again be a consequence of our perturbative expansion, though more likely a result of the droplet ending on a finite temperature bulk horizon.

\begin{figure}
\begin{center}
\includegraphics[scale=.5]{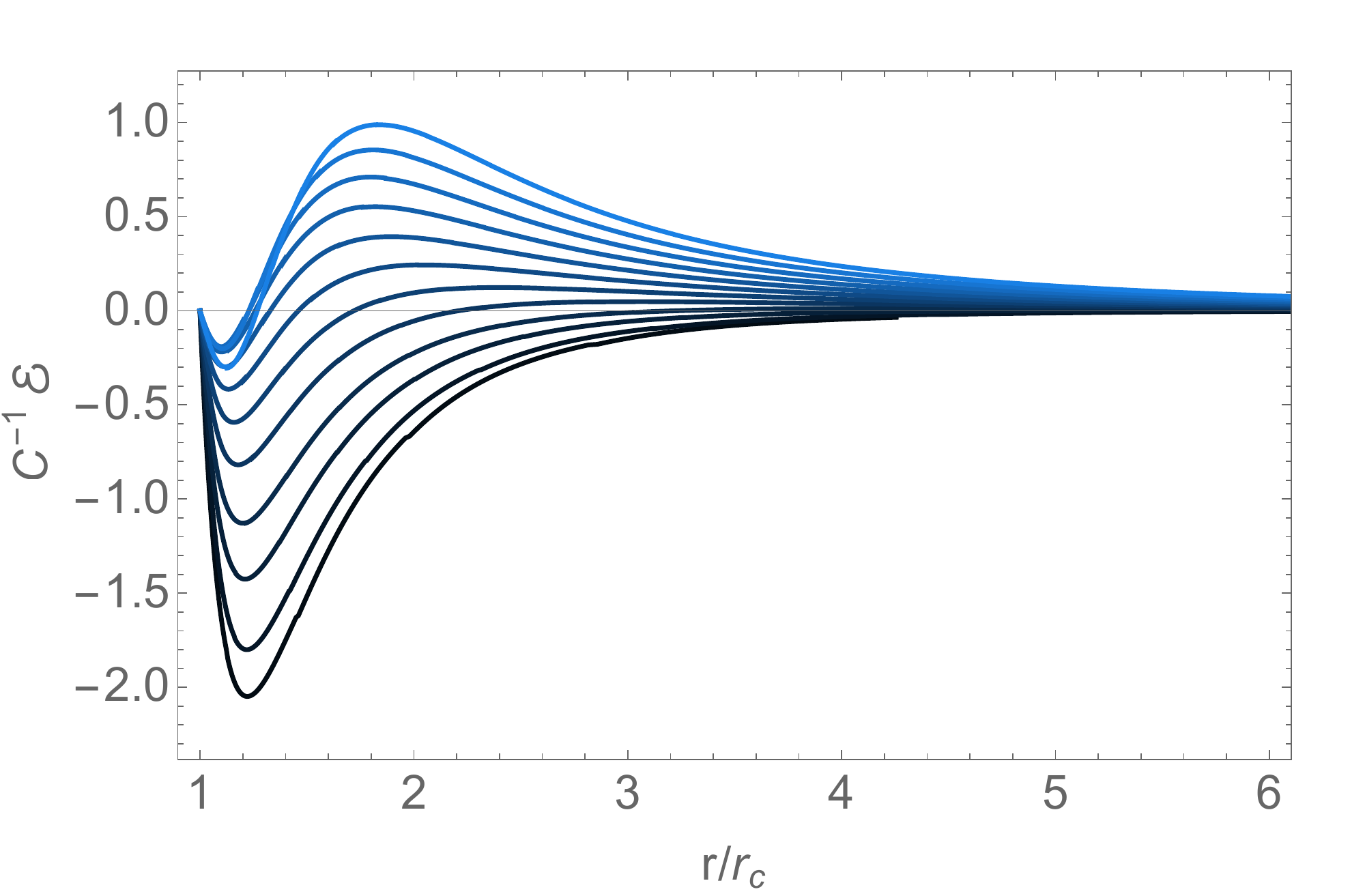}
\caption{\label{energy_new} The energy density seen by an observer with the tangent vector in Eq.~(\ref{newtangent}). Here, we choose $\kappa=.02,\;.05,\;.1,\;.15,\;.2,\;.25,\;.3,\;.35,\;.4,\;.45,\;.5,\;.55$ and $r_c=.1,\; \mu =1, z_c=10\mu^{1/5}$. The color of the curves gets lighter as $\kappa$ increases. For all $\kappa$, the energy density is negative near the outer horizon.}
\end{center}
\end{figure}

To better understand the energy density in regions of the spacetime where $T^a_{\;\;b}$ is not diagonalizable, we instead define a new vector, timelike everywhere outside the outer horizon,
\begin{align}
\label{newtangent}
K= \frac{\partial}{\partial v} + \sqrt{F(z)}\Omega(r,z)\frac{\partial}{\partial \psi}
\end{align}
which also approaches $T_+$ at the outer horizon and goes to $(\partial/\partial v)$ near spatial infinity. An observer with this tangent vector would see the energy density plotted in figure \ref{energy_new}, which is regular everywhere and still has the important feature of being negative near the event horizon. Furthermore, the localization of negative energy density near the event horizon is reminiscent of \cite{Eric2016} and illuminates the ``jammed" nature of the dual CFT. Here too, because $K\to \partial/\partial_v$ near spatial infinity, the energy density also approaches $\mathcal{E}_\infty$, indicative of the CFT in a thermal phase. This tangent vector, however, becomes spacelike inside the outer event horizon, and so is not useful to illustrate strong cosmic censorship. In this region, $t^a$ is well-defined and diverges on the Cauchy horizon.

\begin{figure}
\begin{center}
\includegraphics[scale=.43]{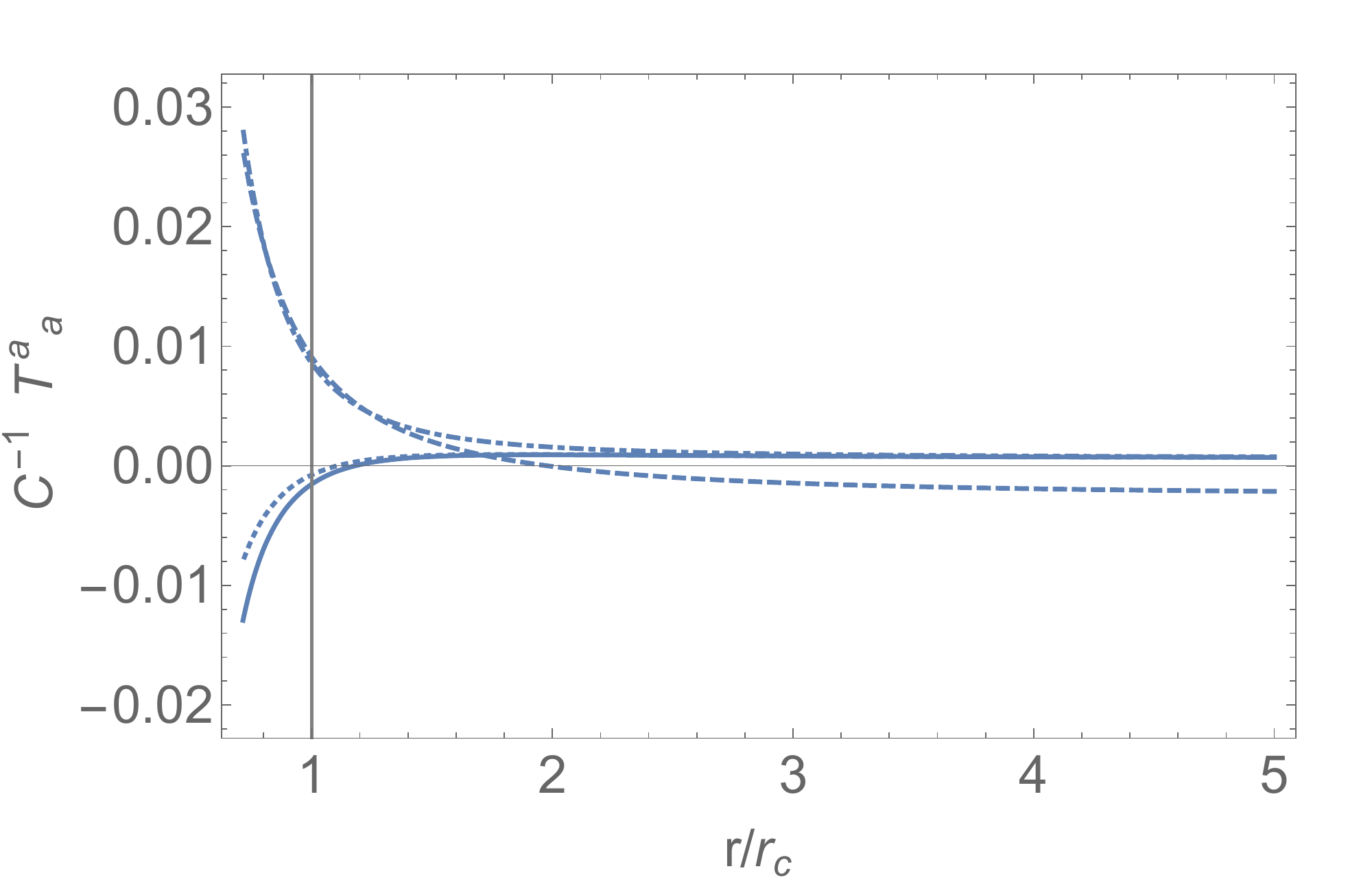}
\includegraphics[scale=.43]{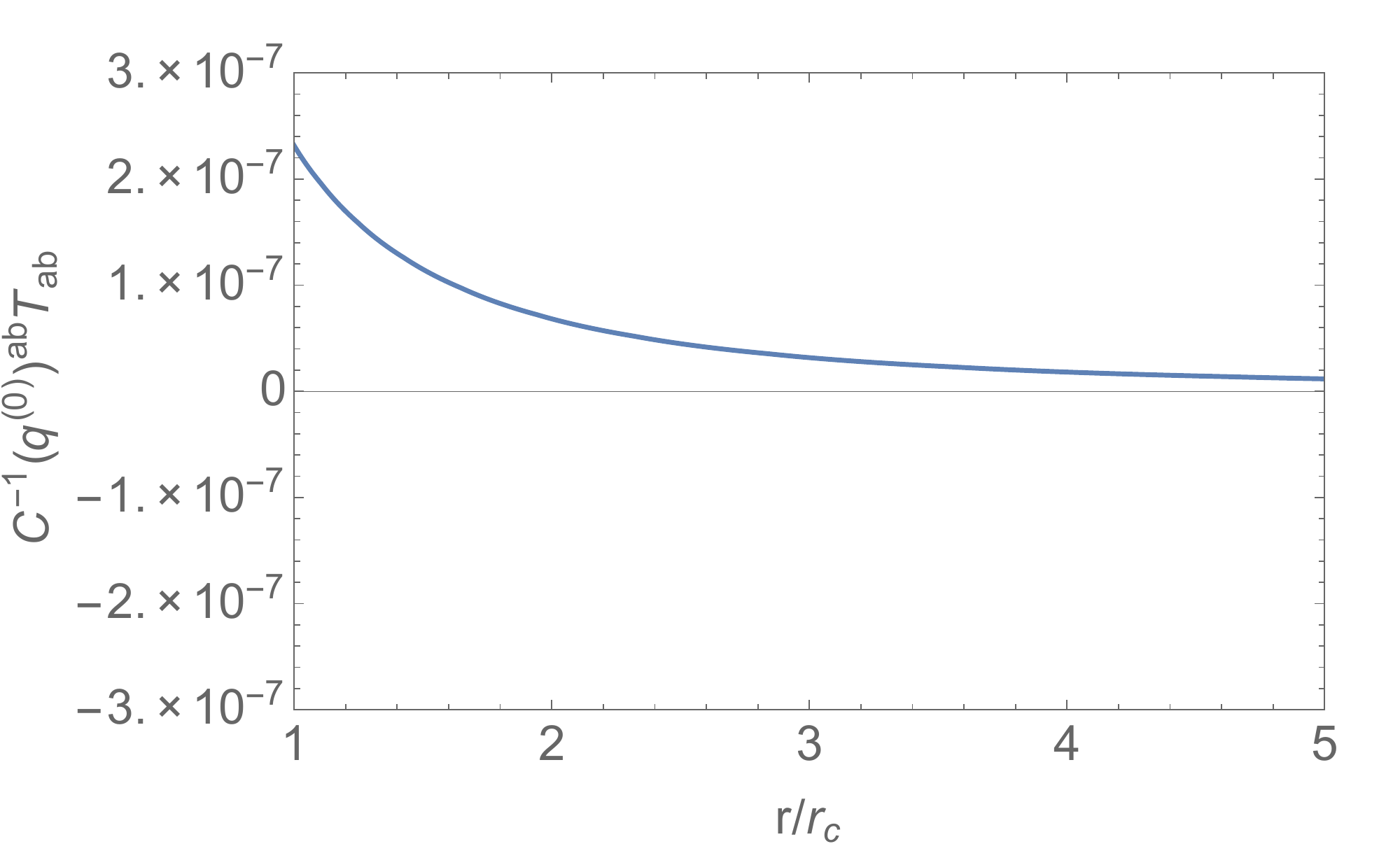}
\caption{\label{diagstress} The left plot displays the diagonal components of the holographic stress-energy tensor, $T^{a}_{\;\;a}$, for $r_c=.1, k=.2, \mu =1, z_c =(10^8\mu)^{1/5}$. The curves correspond to $T^\theta_{\;\;\theta}=T^{\phi}_{\;\;\phi}$ (dotted), $T^v_{\;\;v}$ (dashed), $T^r_{\;\;r}$ (dot-dashed), $T^\psi_{\;\;\psi}$ (thick). Again, the vertical grey line indicates the outer event horizon. The right plot displays the sum of these components. Notably, the trace of the stress-energy tensor vanishes as $\mathcal{O}\left( (\mu/z_c)^{10} \right)$.}
\end{center}
\end{figure}

We emphasize that while the stress-energy tensor diverges on the Cauchy horizon, it is finite at $r=r_c$ so that it is regular on the past and future event horizons (shown in figure \ref{numstress}). Finally, one can check that the trace of stress-energy tensor vanishes at leading order,
 as expected for a CFT in odd spacetime dimensions, just as in the non-rotating case. In figure \ref{diagstress},
 we have plotted the diagonal components of the holographic stress-energy tensor, $C^{-1}( T^{a}_{\;\;a})$ (no sum) as well as the trace. From this figure, 
it is clear that the sum vanishes as we approach the conformal boundary, (i.e. $z_c \to \infty$). One can also check explicitly from the definition of the stress-energy tensor (\ref{energy_momentum1}) 
and the equations of motion for $h_{\mu\nu}$ (\ref{Eq_vr_G}--\ref{Eq_vv_G}), that the trace vanishes as $\mathcal{O}\left((\mu/z_c)^{10}\right)$, exactly following the non-rotating case.

To summarize our numerical results, for generic rotation parameter, $\kappa$, of our boundary black hole, the CFT plasma exhibits the following features. The stress-energy tensor is traceless to leading order in $\mu/z_c$ and regular on the outer event horizon. For a timelike observer, there is a region near the event horizon which has negative energy density. As a timelike observer approaches spatial infinity, the energy density seen by such an observer approaches that of the thermal CFT dual to a six dimensional planar AdS-Schwarzschild black brane. Depending on the observer's velocity, the energy density may remain less than this asymptotic value for all of space, as for the observer with tangent vector $t^a$, or there may be a region with positive energy density, as in the observer with tangent vector $K$. In all cases, this negative energy density diverges on the Cauchy horizon, as shown in Eq.~(\ref{Trr_CH}), in favor of strong cosmic censorship. 

%%%%%%%%%%%%%%%%%%%%%%%%
\section{Conclusion and discussions}
\label{sec:V}
%%%%%%%%%%%%%%%%%%%%%%%%
In this paper we have analytically constructed a rotating black droplet 
solution embedded in the planar Schwarzschild-AdS black brane spacetime 
by applying the generalized derivative expansion method, 
which was originally developed for the static case \cite{Haddad2012}. 
Our method is valid when the horizon size of the black droplet is 
much smaller than the horizon size of the planar Schwarzschild-AdS 
black brane~(and the curvature radius of the background AdS space). 
In this case, the derivative of the metric along the bulk radial direction, $z$, 
is much smaller than the one along the droplet radial direction, $r$
(parallel to the planar horizon). Then, order by order in the derivative 
expansion, we have been able to solve the Einstein equations. 
The horizon radius of the thin black droplet solution gradually shrinks 
toward the planar horizon and caps off smoothly just at the horizon. 
Since the temperature of the black droplet solution is much higher 
than the temperature of the background planar horizon, 
the dual boundary state can be interpreted as 
the Unruh state~\cite{Haddad2013}. 
For our black droplet solution, we have--analytically and holographically--computed the null-null components of the stress-energy tensor for 
a strongly coupled CFT in the boundary five-dimensional rotating 
Myers-Perry black hole spacetime. 
First, we have found that the negative energy appears just outside the event 
horizon, which can be interpreted as a consequence of the particle 
production by the Hawking effect. We show, however, there is no energy flux at infinity, 
as in the static case studied in \cite{Haddad2013}, and therefore 
our boundary CFT can be viewed as a jammed state. 
We have also studied the behavior of the holographic stress-energy tensor 
near the inner Cauchy horizon. The null-null component of the stress-energy 
tensor corresponds to the energy density seen 
by an observer whose world line is transverse to the Cauchy horizon. 
We have found that the null-null component negatively diverges at the Cauchy 
horizon, suggesting that due to quantum effects, the Cauchy horizon would 
become singular, in favor of strong cosmic censorship.

Although we have not analyzed the classical instability 
of our droplet solution in the present paper, we expect our solution to show a
classical instability or divergence of curvature scalars
inside the event horizon. 
In fact, it was shown in \cite{MOK} that in general, adding stationary 
but spatially inhomogeneous linear perturbations makes inhomogeneous 
black branes unstable, rendering the Kretschmann scalar with respect to 
the perturbed geometry divergent on the Cauchy horizon. 
Viewing our black droplet solution as a type of an inhomogeneous 
black string in the bulk and applying the general argument of \cite{MOK}, 
our droplet solution should also exhibit the divergence of curvature scalars  
at the Cauchy horizon even inside the bulk $z<z_c$. 
In the spirit of the bulk-boundary duality, 
our result of the quantum divergence of the stress-energy tensor 
at the Cauchy horizon in the boundary geometry may be viewed 
as a holographic realization of the classical divergence of curvatures  
at the Cauchy horizon in the bulk geometry.

\bigskip 

\noindent 
{\bf Acknowledgments} 

\noindent 
This work was supported in part by
JSPS KAKENHI Grant Number  15K05092(AI), 26400280, 17K05451 (KM) as well as by NSF grant PHY-1504541 (EM). 

\bigskip 

%%%%%%%%%%%%%%%%%%%%%%%%%
\appendix
%%%%%%%%%%%%%%%%%%%%%%%%%
\section{Expressions for ${\cal P}(r)$, ${\cal S}(r)$, ${\cal R}(r)$, and ${\cal Q}(r)$}

We provide the explicite expressions for ${\cal P}(r)$, ${\cal S}(r)$, 
${\cal R}(r)$, and ${\cal Q}(r)$, appeared in (\ref{Eq_vr_G}), (\ref{Eq_gamma_G}), (\ref{Eq_Beta_G}), and 
(\ref{Eq_vv_G}).  
\begin{align}
\label{def P}
& {\cal P}(r)=-2F_c^2\,z_c^{10}(r^4+\kappa^2r_c^4)^3r^2
-5\Biggl[-2r^{14}+r^{12}(r_c^2(1+\kappa^2)+2\sqrt{r^4+\kappa^2r_c^4})
\nonumber \\
&+3\kappa^4r_c^8\,r^4(r_c^2(1+\kappa^2)+2\sqrt{r^4+\kappa^2r_c^4})
+\kappa^6r_c^{12}(r_c^2(1+\kappa^2)+3\sqrt{r^4+\kappa^2r_c^4}) \nonumber \\
&+\kappa^2r_c^{4}\,r^8(3r_c^2(1+\kappa^2)+5\sqrt{r^4+\kappa^2r_c^4})
-r_c^{2}\,r^{10}(6r_c^2\kappa^2+(1+\kappa^2)\sqrt{r^4+\kappa^2r_c^4}) \nonumber \\
&-r_c^{6}\,r^{6}\kappa^2(6r_c^2\kappa^2+(1+\kappa^2)\sqrt{r^4+\kappa^2r_c^4})
-2r_c^{10}\,\kappa^4r^{2}(r_c^2\kappa^2+3(1+\kappa^2)\sqrt{r^4+\kappa^2r_c^4})\Biggr]\mu^{10}
\nonumber \\
&+2F_c\,z_c^{5}(r^4+\kappa^2r_c^4)\Biggl[-6r^8 \sqrt{r^4+\kappa^2r_c^4}\,\mu^5
-9r^4\kappa^2r_c^4 \sqrt{r^4+\kappa^2r_c^4}\,\mu^5-3r_c^8\,\kappa^4 \sqrt{r^4+\kappa^2r_c^4}\,\mu^5
\nonumber \\
&+r^{10}(z_c^5+5\mu^5)+r^2r_c^6\,\kappa^2\Bigl\{6(1+\kappa^2)\sqrt{r^4+\kappa^2r_c^4}\,\mu^5+r_c^2\,\kappa^2(z_c^5+5\mu^5)\Bigr\}
\nonumber \\
&+r^6r_c^2\Bigl\{3(1+\kappa^2)\sqrt{r^4+\kappa^2r_c^4}\,\mu^5+2r_c^2\,\kappa^2(z_c^5+5\mu^5)\Bigr\}\Biggr], 
\end{align}

\begin{align}
\label{def_S}
& {\cal S}(r)=\frac{10\kappa\sqrt{1+\kappa^2}r_c^3\,r^2}{z_c^6}\Biggl[2F_c^2z_c^{10}r^2(r^4+\kappa^2r_c^4)^3+5\Bigl\{
-3r^{14}+2(1+\kappa^2)r_c^2\,r^{12}+2\kappa^4(1+\kappa^2)r_c^{10}\,r^4 \nonumber \\
&-\kappa^6\sqrt{r^4+\kappa^2r_c^4}\,r_c^{12}+\kappa^2r_c^4(4(1+\kappa^2)r_c^2+5\sqrt{r^4+\kappa^2r_c^4})r^8
+r_c^2(-5r_c^2\kappa^2+(1+\kappa^2)\sqrt{r^4+\kappa^2r_c^4})r^{10} \nonumber \\
&+r_c^{10}\,\kappa^4(r_c^2\kappa^2+2(1+\kappa^2)\sqrt{r^4+\kappa^2r_c^4})r^{2}
-r_c^6\,\kappa^2(r_c^2\kappa^2+5(1+\kappa^2)\sqrt{r^4+\kappa^2r_c^4})r^6\Bigr\}\mu^{10} \nonumber \\
&+2F_c z_c^5\Bigl\{6\sqrt{r^4+\kappa^2r_c^4}\,\mu^5r^{12}+7r_c^4\kappa^2\sqrt{r^4+\kappa^2r_c^4}\,\mu^5r^8
-r_c^{12}\kappa^6\sqrt{r^4+\kappa^2r_c^4}\,\mu^5-(z_c^5+5\mu^5)r^{14} \nonumber \\
&+r_c^{10}\kappa^4(2(1+\kappa^2)\sqrt{r^4+\kappa^2r_c^4}\,\mu^5-r_c^2\kappa^2(z_c^5-3\mu^5))r^2 \nonumber \\
&+r_c^6\kappa^2(-5(1+\kappa^2)\sqrt{r^4+\kappa^2r_c^4}\,\mu^5-r_c^2\kappa^2(3z_c^5-\mu^5))r^6 \nonumber \\
&-r_c^2(3(1+\kappa^2)\sqrt{r^4+\kappa^2r_c^4}\,\mu^5+r_c^2\kappa^2(3z_c^5+7\mu^5))r^{10}\Bigr\}\Biggr],  
\end{align}
\begin{align}
\label{def_R}
& {\cal R}(r)=\frac{10\kappa^2 r_c^4\,r}{z_c^6}\Biggl[2F_c^2z_c^{10}\,r^2(r^4+\kappa^2r_c^4)^2(5r^4+\kappa^2r_c^4)
+5\Bigl\{4r_c^{10}\,\kappa^4(1+\kappa^2)r^4-\kappa^6r_c^{12}\sqrt{r^4+\kappa^2r_c^4} \nonumber \\
&+5r_c^2(1+\kappa^2)\sqrt{r^4+\kappa^2r_c^4}\,r^{10}
-5r_c^6\kappa^2(1+\kappa^2)\sqrt{r^4+\kappa^2r_c^4}\,r^6+2r_c^{10}\kappa^4(1+\kappa^2)\sqrt{r^4+\kappa^2r_c^4}\,r^2
\nonumber \\
&+(4r_c^2(1+\kappa^2)-6\sqrt{r^4+\kappa^2r_c^4})r^{12}
+\kappa^2r_c^4(8r_c^2(1+\kappa^2)+3\sqrt{r^4+\kappa^2r_c^4})r^8\Bigr\}\mu^{10} \nonumber \\
&-2F_cz_c^5\Bigl\{-6\sqrt{r^4+\kappa^2r_c^4}\,\mu^5r^{12}-7r_c^4\kappa^2\sqrt{r^4+\kappa^2r_c^4}\,\mu^5\,r^8
+r_c^{12}\kappa^6\sqrt{r^4+\kappa^2r_c^4}\,\mu^5+(5z_c^5+\mu^5)r^{14} \nonumber \\
&+\kappa^2r_c^6\{5(1+\kappa^2)\sqrt{r^4+\kappa^2r_c^4}\,\mu^5+r_c^2\kappa^2(7z_c^5-5\mu^5)\}r^6
\nonumber \\
&+\kappa^4r_c^{10}\{-2(1+\kappa^2)\sqrt{r^4+\kappa^2r_c^4}\,\mu^5+r_c^2\kappa^2(z_c^5-3\mu^5)\}r^2
\nonumber \\
&+r_c^2\{3(1+\kappa^2)\sqrt{r^4+\kappa^2r_c^4}\,\mu^5+r_c^2\kappa^2(11z_c^5-\mu^5)\}r^{10}\Bigr\}\Biggr], 
\end{align}
\begin{align}
\label{def_Q}
& {\cal Q}(r)=\frac{5\mu^5r^2}{z_c^6}\Biggl[6(2z_c^5+3\mu^5)r^{16}-2\kappa^6(1+\kappa^2)r_c^{14}(2z_c^5+3\mu^5)\sqrt{r^4+\kappa^2r_c^4}
\nonumber \\
&-r_c^8\,\kappa^2(1+\kappa^2)\Bigl\{10(1+\kappa^2)(2z_c^5+3\mu^5)\sqrt{r^4+\kappa^2r_c^4}+r_c^2\,\kappa^2(8z_c^5+7\mu^5)\Bigr\}r^6
\nonumber \\
&-3\Bigl\{2(2z_c^5+3\mu^5)\sqrt{r^4+\kappa^2r_c^4}+r_c^2\,(1+\kappa^2)(8z_c^5+7\mu^5)\Bigr\}r^{14} \nonumber \\
&+\kappa^4r_c^{12}\Bigl\{(4+9\kappa^2+4\kappa^4)(2z_c^5+3\mu^5)\sqrt{r^4+\kappa^2r_c^4}+r_c^2\,\kappa^2(1+\kappa^2)(8z_c^5+7\mu^5)\Bigr\}r^2
\nonumber \\
&-r_c^4\Bigl\{5r_c^2\,\kappa^2(1+\kappa^2)(8z_c^5+7\mu^5)+\sqrt{r^4+\kappa^2r_c^4}\,\{2z_c^5(6+19\kappa^2+6\kappa^4)
+(-2+17\kappa^2-2\kappa^4)\mu^5\}\Bigr\}r^{10}
\nonumber \\
&+5r_c^2\Bigl\{3(1+\kappa^2)(2z_c^5+\mu^5)\sqrt{r^4+\kappa^2r_c^4}+2r_c^2(2z_c^5\kappa^2+(1+5\kappa^2+\kappa^4)\mu^5) \Bigr\}r^{12}
\nonumber \\
&-2r_c^{10}\kappa^4\Bigl\{(1+\kappa^2)(2z_c^5+3\mu^5)\sqrt{r^4+\kappa^2r_c^4}+r_c^2(2z_c^5\kappa^2-(5+7\kappa^2+5\kappa^4)\mu^5) \Bigr\}r^4
\nonumber \\
&+r_c^6\kappa^2\Bigl\{(1+\kappa^2)(38z_c^5+47\mu^5)\sqrt{r^4+\kappa^2r_c^4}
+2r_c^2(2z_c^5\kappa^2+(10+23\kappa^2+10\kappa^4)\mu^5) \Bigr\}r^8\Biggr]. 
\end{align}

%%%%%%%%%%%%%%%%%%%%%%%%%%%%%%%%%%
%\begin{figure}
% \begin{center}
%  \includegraphics[width=7truecm,clip]{DC_conductivity_velocity.eps}
%  \caption{(color online) {\color{red}$\eta=L^4{\cal G}/(\epsilon^2 r_+^4)$} 
%is plotted for $k=1/2$ for various $\beta$. 
%$\beta=0.1$, $\beta=0.07$, and $\beta=0.05$ correspond to a circle, a rhombus, and 
%a square, respectively.} 
% \end{center}
%\end{figure}
%%%%%%%%%%%%%%%%%%%%%%%%%%%%%%%%%%%


\begin{thebibliography}{99} 
\bibitem{Haddad2012}
N.~Haddad, ``Black Strings Ending on Horizons," 
Class. Quantum Grav. {\bf 29} (2012) 245001. 
\bibitem{EHNO2009}
R.~Emparan, T.~Harmark, V.~Niarchos, and N.~A.~Obers, 
Phys.~Rev.~Lett.~{\bf 102} (2009) 191301. 
\bibitem{Armas2013}
J.~Armas, JHEP {\bf 1309} (2013) 073
\bibitem{FPR1998}
L.~H.~Ford, M.~J.~Pfenning, and T.~A.~Roman, 
``Quantum inequalities and singular negative energy densities'', 
Phys.~Rev.~{\bf D 57} (1998) 4839.  
\bibitem{Hiscock1977}
W.~A.~Hiscock, 
``Stress-energy tensor near a charged, rotating, evaporating black hole,'' 
Phys.~Rev.~{\bf D 15} (1977) 3054. 
\bibitem{HiscockKonkowski1982}
W.~A.~Hiscock and D.~A.~Konkowski, 
``Quantum vacuum energy in Taub-NUT (Newman-Unti-Tamburino)-type cosmologies,'' 
Phys.~Rev.~{\bf D 26} (1982) 1225. 
\bibitem{adscft}
J. Maldacena, 
``{The Large N limit of superconformal field theories and supergravity}'' 
Adv. Theor. Math. Phys. {\bf 2} (1998) 231,   
O. Aharony, S. Gubser, J. Maldacena, H. Ooguri, and Y. Oz, 
``{Large N field theories, string theory and gravity}''
Phys. Rep. {\bf 323} (2000) 183. 
\bibitem{HMR2010-1}
V.~E.~Hubeny, D.~Marolf, and M.~Rangamani, 
``Hawking radiation in large N strongly-coupled field theories," 
Class.~Quantum~Grav. {\bf 27} (2010) 095015. 
\bibitem{HMR2010-2}
V.~E.~Hubeny, D.~Marolf, and M.~Rangamani, 
``Hawking radiation from AdS black holes," 
Class.~Quantum~Grav. {\bf 27} (2010) 095018. 
\bibitem{HMR2010-3}
V.~E.~Hubeny, D.~Marolf, and M.~Rangamani, 
``Black funnels and droplets from the AdS C-metrics," 
Class.~Quantum~Grav. {\bf 27} (2010) 025001.
\bibitem{CDMS2011}
M.~M.~Caldarelli, O.~J.~C.~Dias, R.~Monteiro, and J.~E.~Santos, 
``Black funnels and droplets in thermal equilibrium," 
JHEP {\bf 1105} (2011) 116. 
\bibitem{FMS2012}
S.~Fischetti, D.~Marolf, and J.~E.~Santos, 
``AdS flowing black funnels: Stationary AdS black holes with non-Killing horizons 
and heat transport in the dual CFT,'' 
Class.~Quantum~Grav. {\bf 30} (2013) 075001. 
\bibitem{FischettiMarolf2012}
S.~Fischetti and D.~Marolf,
``Flowing Funnels: Heat sources for field theories and the AdS$_3$ dual of CFT$_2$ Hawking radiation,'' 
Class.~Quantum~Grav. {\bf 29} (2012) 105004. 
\bibitem{SantosWay2012}
J.~E.~Santos and B.~Way, ``Black Funnels," 
JHEP {\bf 1212} (2012) 060. 
\bibitem{Haddad2013}
N.~Haddad, ``Hawking Radiation from Small Black Holes at Strong Coupling and Large N," 
Class.~Quantum~Grav. {\bf 30} (2013) 195002.   
\bibitem{FischettiSantos2013}
S.~Fischetti and J.~E.~Santos, ``Rotating Black Droplet," 
JHEP {\bf 1307} (2013) 156. 
\bibitem{FiguerasTunyasuvunakool2013}
P.~Figueras and S.~Tunyasuvunakool 
``CFTs in rotating black hole backgrounds,"   
Class. Quantum Grav. {\bf 30} (2013) 125015.
\bibitem{Eric2016}
E.~Mefford, ``Entanglement Entropy in Jammed CFTs," 
arXiv:1605.09369~[hep-th].     
\bibitem{MyersPerry}
R.~C.~Myers and M.~Perry, ``Black holes in higher Dimensional Space-Times," 
Annals Phys {\bf 172} (1986) 304.  
\bibitem{EJM1999}
R.~Emparan, C.~V.~Johnson, and R.~C.~Myers, 
``Surface Terms as Counterterms in the AdS/CFT Correspondence," 
Phys.~Rev.~{\bf D60} (1999) 104001.  
\bibitem{SantosWay2014} 
  J.~E.~Santos and B.~Way,
  ``Black Droplets,''
  JHEP {\bf 1408}, 072 (2014).
\bibitem{Figueras2011} 
  P.~Figueras, J.~Lucietti and T.~Wiseman,
  ``Ricci solitons, Ricci flow, and strongly coupled CFT in the Schwarzschild Unruh or Boulware vacua,''
  Class.\ Quant.\ Grav.\  {\bf 28}, 215018 (2011).
  
\bibitem{MOK} 
K.~Maeda, T.~Okamura, and J.I.~Koga, 
``Inhomogeneous charged black hole solutions in asymptotically anti-de Sitter spacetime," Phys.~Rev.~D{\bf 85} (2012) 066003. 
\end{thebibliography}
\end{document}